\documentclass{aa}
\usepackage{times,graphics}
\begin{document}
\newcommand{\bs}{\boldmath}
\thesaurus{11(11.11.1; 11.16.1; 11.19.6; 08.11.1; 09.11.1)}
\title{Mass distribution and kinematics of the barred galaxy NGC\,2336}
\author{K.\, Wilke\inst{1}\fnmsep\thanks{Visiting astronomer of the
German-Spanish Astronomical Center, Calar Alto, operated by the
Max-Planck-Institut f\"ur Astronomie, Heidelberg jointly with the
Spanish National Comission for Astronomy}
\and C.\,M\"ollenhoff\inst{1}\fnmsep$^\star$ \and
M.\,Matthias\inst{2}}
\institute{Landessternwarte Heidelberg, K\"onigstuhl, D-69117
Heidelberg, Germany
\and Astronomisches Institut, Universit\"at Basel, Venusstr.\,7,
CH-4102 Binningen, Switzerland}
\offprints{kwilke@lsw.uni-heidelberg.de}
\date{Received / Accepted}
\maketitle

\begin{abstract}
For the intermediate-type barred galaxy NGC\,2336 stationary models
are constructed which reproduce in a consistent manner the observed
distribution of the luminous matter and the observed gas kinematics in
those regions affected by the bar. We present 2D fits to the
observed $NIR$-band luminosity distribution that consist of three
components: a bulge, a bar, and a disk. The brightness distribution
of each component is converted into an underlying mass distribution
by means of a suitable $M/L$-conversion. The resulting coadded
potential of NGC\,2336 is implemented into a numerical code for the
computation of closed orbits for gas clouds (HII-gas). Using the
resulting tracks, the phase space accessible to the models is
examined with regard to the main orbit families. For different orbit
energies complete sets of closed orbits are computed. By projection
to the reference frame of the galaxy, artificial rotation curves for
every model are obtained and are compared with the observed rotation
curves of the HII-gas.

In an iterative procedure, the parameters of the NGC\,2336-models are
optimized by computing and evaluating a large number of parameters. 
The result is a final model that reproduces the morphological
structure of NGC\,2336 as well as the observed kinematics of the HII-gas.
The parameter values from the morphological decomposition and
those needed to fit the HII-rotation curves best are in exellent
agreement.  The effects of changing single parameter values and
possible error sources are discussed in detail. It turns out that the
kinematics of the warm HII-gas of NGC\,2336 can be explained
without considering hydrodynamic effects, even in the central regions.
\end{abstract}

\section{Introduction}
Due to their strong non-axisymmetric potential, barred galaxies
exhibit peculiar stellar and especially peculiar gas kinematics when
compared to normal disk galaxies without bar.  Since deconvolving
their non-axisymmetric velocity field is difficult, the computation of
consistent multi-component stationary models for these objects has
proven to be a suitable method to gain further insight into the
internal mass distribution and kinematics.  This stationary modelling
procedure can answer the question whether the observed kinematics of
the galaxy agree with the model predictions derived from the observed
distribution of luminous matter (assumed that plausible corrections
for the presence of dark matter can be applied).

For the following studies the northern hemisphere object NGC\,2336 was
chosen. NGC\,2336 is an intermediate-type spiral of Hubble-type
$SAB(r)bc$ with a prominent bar. This galaxy has not been subject to
extensive morphological and kinematical studies yet. Basic parameters
given by Tully (\cite{tully}) are listed in Table\ref{baspar1}.
NGC\,2336 contains sufficient amounts of warm HII-gas for
emission-line-spectroscopy. As is also required for kinematic studies
and deprojection, the bar lies in a non-degenerate position with
respect to line of sight, i.e. its apparent major axis
does not coincide with the kinematic line of nodes (denoted $LON$
hereafter).
The
inclination of $i=59^{\circ}$ ensures a fairly precise determination
of the apparent axes ratios of disk, bulge and bar.

\begin{table}[h]
\begin{center}
\begin{tabular}{|lc|}\hline
$\alpha_{2000}$ & $07^h\,18^m\,00^s$\\
$\delta_{2000}$ & $80^{\circ}\,16'\,00''$\\
$i$ & $59^{\circ}$\\
$B_T^{b,i}$ [mag] & 10.65\\
$D_{25}^{b,i}$ [arcmin]& $5.2$\\
$R$ [Mpc]& 22.9\\
[1.0ex]
$L_B$ [$L_{\sun}$] &$9.77\cdot 10^{10}$\\
$M_T$ [$M_{\sun}$] &$3.63\cdot 10^{11}$\\
$M_{HI}/M_T$ & 0.25\\
$M_T/L_B$ & 3.7\\
\hline
\end{tabular}
\end{center}
\caption{\label{baspar1}\em Basic parameters of NGC\,2336 as given in
Tully (\cite{tully}). The
quantities are defined as follows:
$\alpha_{2000},\,\beta_{2000}$:
right
ascension and declination at the epoch 2000.0,
$i$: inclination from face-on,
$B_T^{b,i}$: blue apparent
magnitude corrected for reddening due to internal and external
absorption, $D_{25}^{b,i}$: diameter at the
$25\hbox{mag}/\hbox{arcsec}^2$ blue isophote, adjusted for projection and
obscuration effects,
$R$: distance based on $H_0=100$km/sec,
$L_B$: intrinsic blue luminosity,
$M_T$: total mass of NGC\,2336,
$M_{HI}/M_T$: ratio of $HI$-mass
to blue luminosity,
$M_T/L_B$ ratio of total mass to
blue luminosity.
}
\end{table}

Former photometric studies of NGC\,2336 were mainly restricted to the
HI observations of van Moorsel (\cite{moorsel}).  In HI, this galaxy
reveals a very regular morphological structure devoid of any kind of
anomalies or major asymmetries, except for the nearly complete lack
of HI in the central regions. The HI distribution supports the image of a
mainly undisturbed spiral structure in the outer part of the disk
with numerous star forming regions. Although NGC\,2336 belongs to an
apparent pair of galaxies (together with IC\,467) with a projected
linear distance of $135$kpc, its undisturbed disk does not exhibit any
distinct sign of recent interactions. No
morphological structures connecting these objects have been found. 
Due to the large beam size used
for the HI-observations ($d=35''$), the inner part of the disk with
the bar is not resolved.  The HI disk isophotes at large radii remain
undisturbed down to the detection limit. With lower threshold
values of $n_{HI}\ge 1.2\cdot 10^{20}$atoms/$\hbox{cm}^2$, the disk
extends up to $\approx 55$kpc ($8.2'$) from the center. Recent
studies by Martin (\cite{martin}) based on POSS plates yielded the
basic morphological parameters for the bar listed in Table
\ref{barpar1}.

According to the results of the HI-studies by van Moorsel
(\cite{moorsel}), NGC\,2336 does not exhibit any kinematic
peculiarities up to a distance of $5'$ ($33$kpc) from the center,
i.e., the HI-velocity field is typical for a disk-dominated 
galaxy.  A distortion of the velocity field at $40''$ north to the
center is caused by a sudden drop of the HI-column density and does
not indicate non-circular motions in that region. The velocity field
shows that the southern part of the galaxy is inclined towards us.

\begin{table}[h]
\begin{center}
\begin{tabular}{|ccccccc|}\hline
$PA_{disk}$&$a_{bar}$&$b_{bar}$&$b/a_{bar}$&$L_b$&$b/a(i)$&$L_b(i)$\\[0.4ex]
$178^{\circ}$&$20''$&$14''$&$0.70$&$0.09$&$0.59$&$0.17$\\ \hline
\end{tabular}
\end{center}
\caption{\label{barpar1}\em Morphological parameters for
NGC\,2336. The results are based on POSS-plate measurements
by Martin (\cite{martin}). Parameters are:
$PA_{disk}$: position angle of the
apparent disk major axis,
$a_{bar},\,b_{bar}$: major and
minor axis of the bar,
$b/a_{bar},\,L_b$: apparent bar axis ratio
and relative bar length compared to $D_{25}^{b,i}$,
$b/a(i),\,L_b(i)$: the same as
above, but adjusted for deprojection.
}
\end{table}

Because of the low resolution and the lack of HI just in the inner
region, the HI velocity field is not suited to study the
perturbation of the velocity field due to the bar potential.
Optical spectrograms which show the kinematics of stars and (warm) gas
with a much better resolution have not been available up to now.

The aim of this paper is the quantitative understanding and modelling
of the bar-perturbed velocity field in NGC\,2336.  For that purpose we
study the NIR morphology of this galaxy, representing the ditribution
of the luminous matter.  The total potential of disk, bar, and bulge
is calculated. We consider the HII-gas to be in a stationary motion in this
potential. The subsequent kinematical modelling leads to artificial rotation
curves which are compared with the observed velocity field. The optimal
parameters for disk, bar, and bulge, as well as the intrinsic geometry of these
components, are obtained by
an iterative study of many different models and their closed orbits.

The paper is divided into the following parts: Section 2 addresses the
photometric observations, section 3 introduces the procedure for
constructing multi-component 2D mass models and their subsequent
deprojection.  Section 4 presents the data reduction process and the
detailed results of the spectroscopic observations of both, the
gaseous and the stellar component of NGC\,2336. A basic outline of the
numerical representation of the potentials used for the orbit
integrator (which computes closed orbits by means of a numerical
FORTRAN-/C-program) is given in Section 5. Section 6 refers to the
different Lindblad resonances in the NGC\,2336-models, to the
different orbit families and their relative contribution to the total
number of orbits in a model.  Model units are discussed in Section 7,
while the basic instrument for phase space analysis -- the {\em
Poincar\'e}-Surfaces of Section -- are introduced in Section 8 where the
characteristic features of the NGC\,2336-phase space are discussed.
Finally, section 9 deals with the variation of independent model
parameters which lead to the optimal model solution in Section
10. Results are discussed in Section 11.

\section{Surface Photometry}
\subsection{NIR Images}
To avoid wrong estimates for the relative contribution of the model
components of NGC\,2336 to the total luminosity, all models are based
on NIR images ($J$-band) where dust absorption is much less
important than in the optical wavelength range. Additionally, galactic
structures appear much more regular due to the smoother distribution
of cool giant stars whose light is mainly traced in the NIR (for a
review see Frogel et al.~\cite{frogel}).

The images used for our studies were obtained using the
$MAGIC$-camera at the MPIA $2.2m$-telescope at Calar Alto (Spain)
during September 1995. $MAGIC$ consists of a $256\times 256$ NICMOS-chip
and provides a field of view of $171''$ with a resolution of
$0.67''$/pixel when mounted at the $2.2m$-telescope ($f/8$). Images
were obtained in the $J$- ($1.2\mu$m) and $K$-band ($2.2\mu$m).

By repeating a routine which centers the telescope alternately on the
object and neighbouring sky fields, 48 object exposures and skyflats
with exposure times of $t_{exp}=10$sec were obtained, resulting in a
total exposure time of $t_{exp}=480$sec. For the skyflats, the
telescope was moved 25' away from the object field in several
directions. An additional telescope offset of a few pixels between the
repetitions of the routine was also applied to correct for bad
pixels on the detector chip.  Several domeflat series were exposed to
correct for illumination effects on the NICMOS array.

\subsection{Data Reduction}
\begin{figure*}
\resizebox{\hsize}{!}{\includegraphics{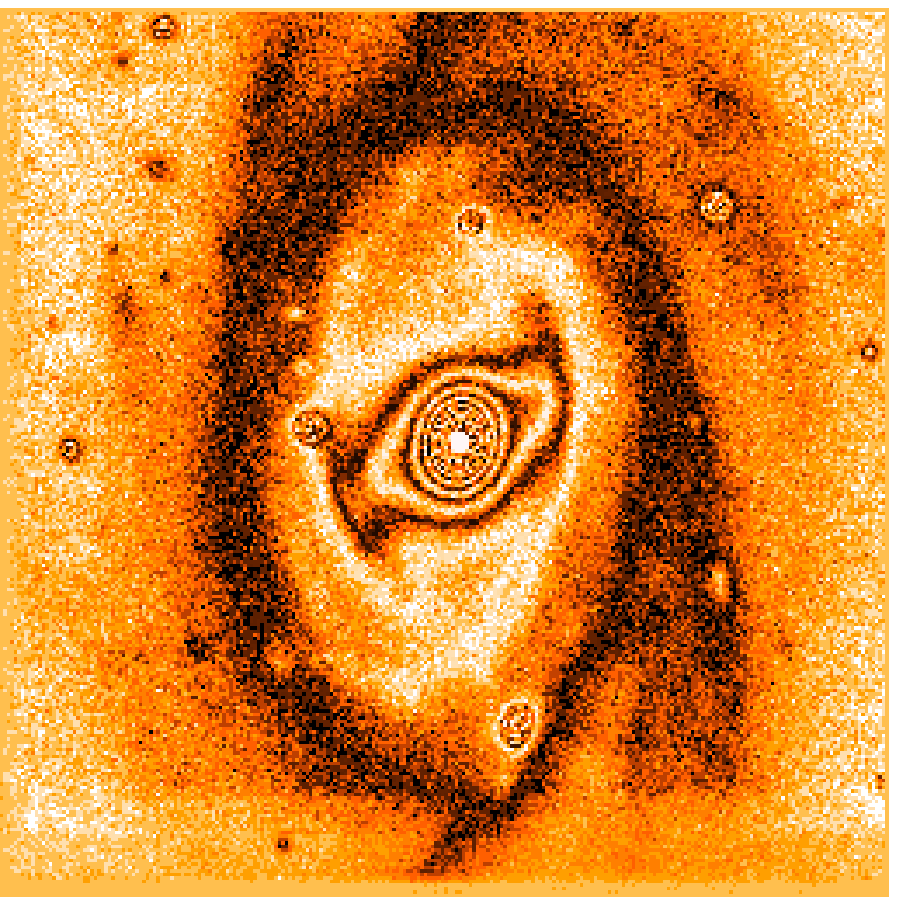}\hspace{0.2cm}
\includegraphics{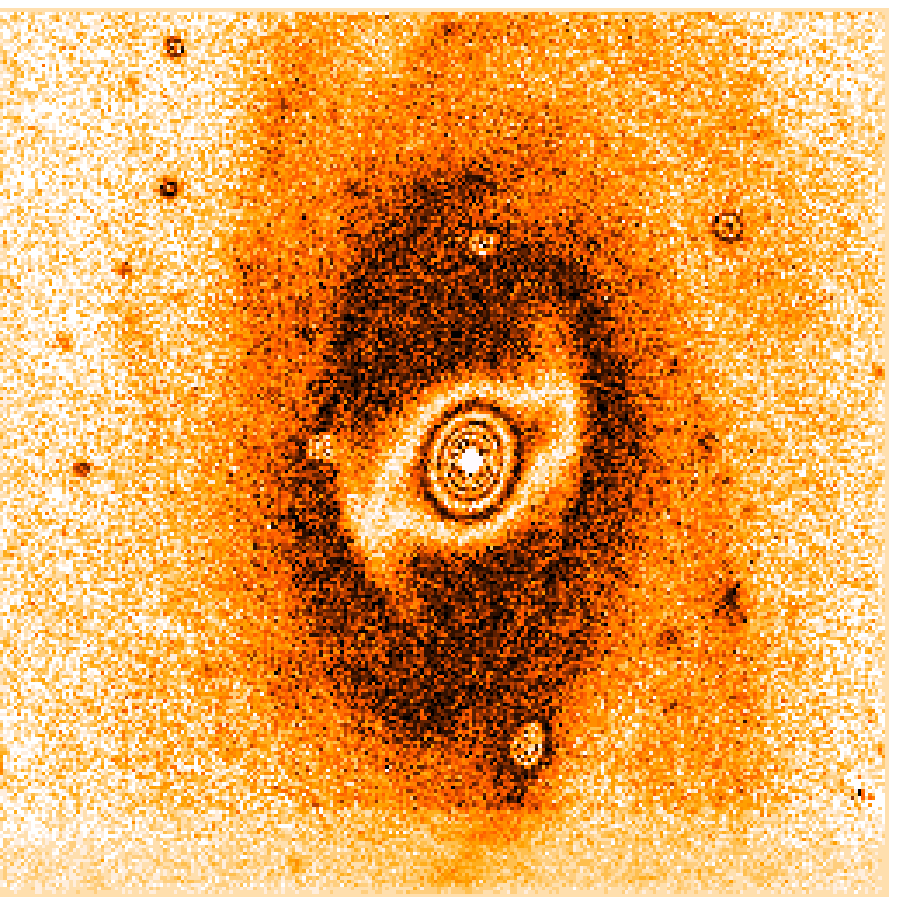}}
\caption{\em $J$- (left) and $K$-band image (right) of
NGC\,2336 with a resolution of $0.67''$/pixel. The disk extends well
beyond the area of the NICMOS-detector. Due to the stronger airglow
emission, the $S/N$-ratio in $K$ is 
lower than in $J$.}
\label{2336jk}
\end{figure*}

The single exposures of NGC\,2336 were reduced by a standard reduction
process: A 2D sky pattern is constructed by using a vertical median
filter for the sky fields that were obtained in different offset
directions. The resulting  image is normalized to zero.  The sky
contribution is subtracted by linear interpolation of the sky level
between the sky exposure before and after each object frame. By
median-filtering the domeflat series, a master domeflat is created and
normalized to one.  The 2D sky pattern is subtracted from the object
frame and the result is divided by the master domeflat.  The
pre-reduced object frames are then centered on field stars and then
co-added by a $\kappa$-$\sigma$-clipping algorithm to avoid the
disadvantages of a simple addition (bad pixels remain) or a simple
median filtering (quantitative distortion of the images, lower
$S/N$-ratio).  The resulting images of NGC\,2336 are shown in
Fig.~\ref{2336jk}.

\section{Modelling the Mass Distribution}
\begin{figure*}[t]
\resizebox{\hsize}{!}{
\includegraphics{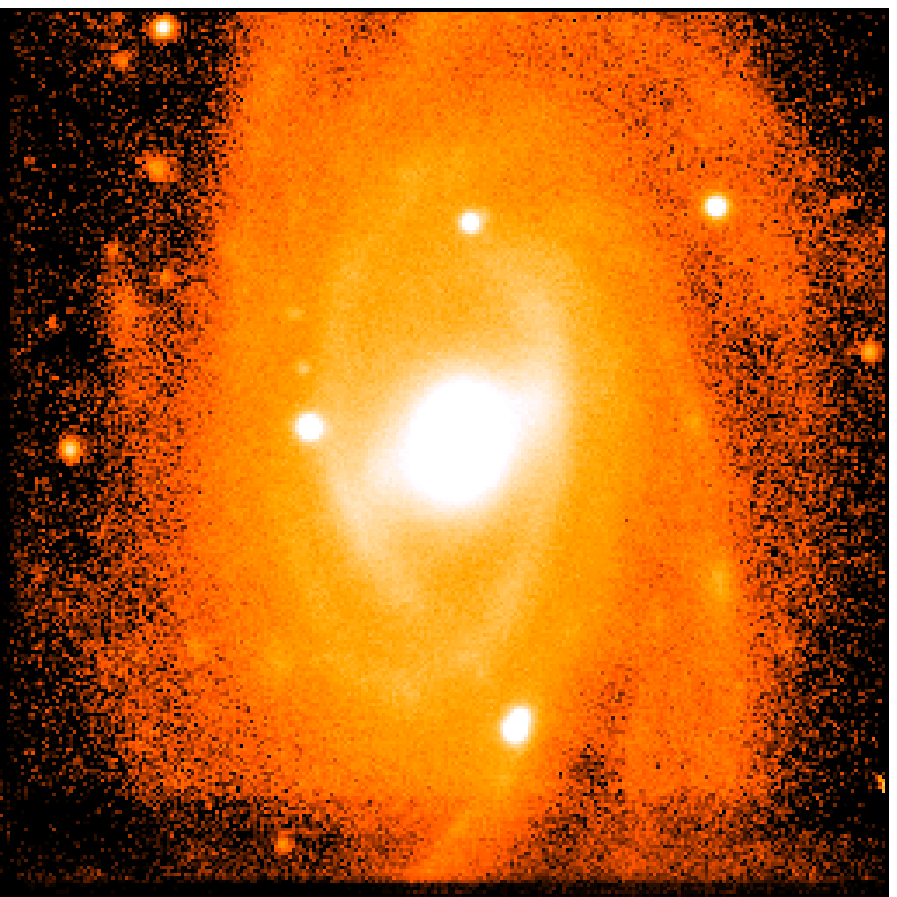}\hspace{0.2cm}
\includegraphics{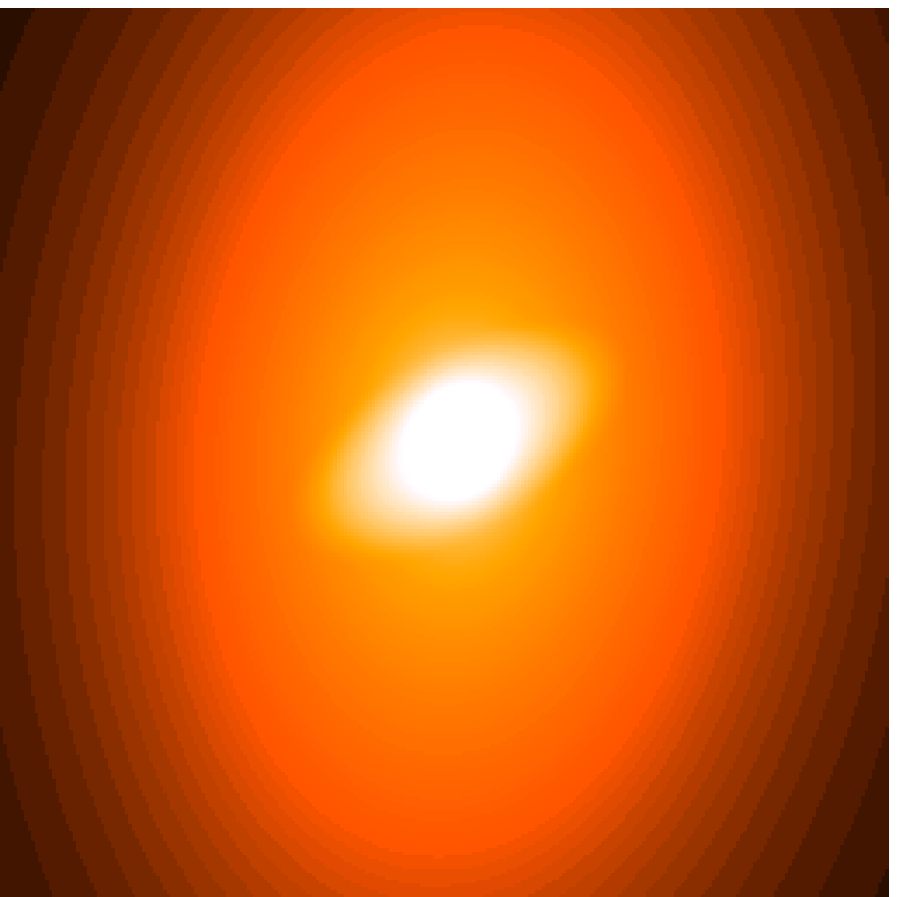}\hspace{0.2cm}
\includegraphics{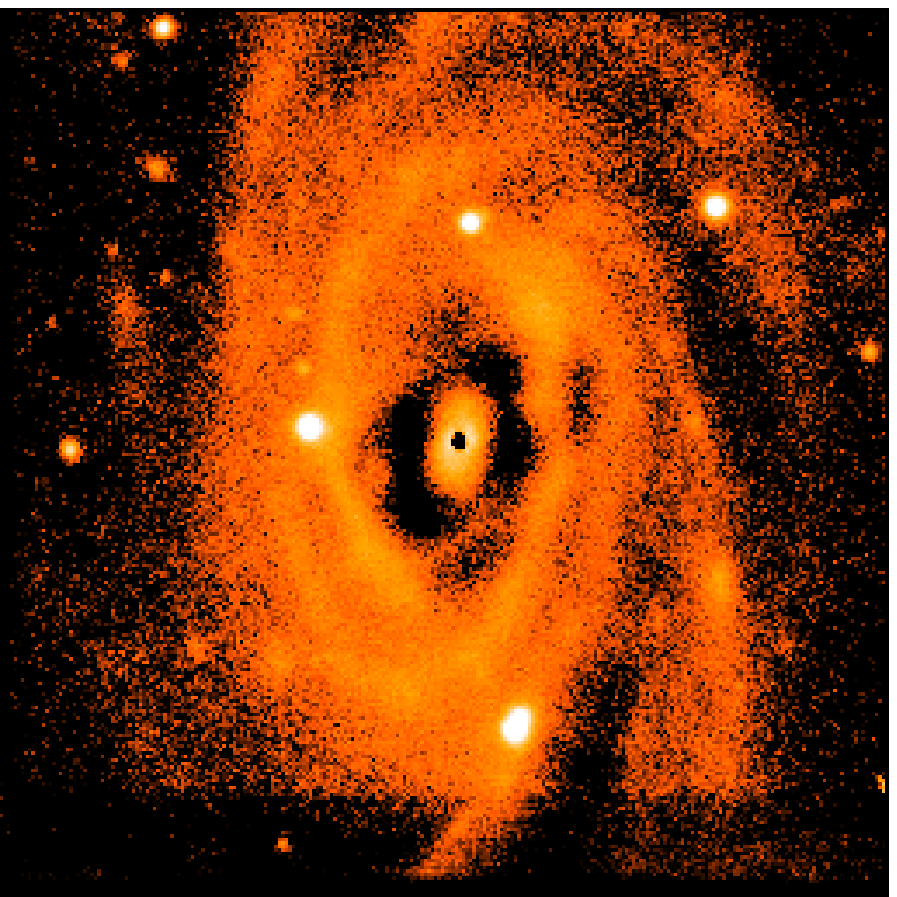}}
\caption{\em
Left: $J$-band image of NGC\,2336 ($0.67''$/pixel), center:
2D three-component model of the observed luminosity distribution,
right: residuals remaining after subtraction of the model from the
original image.}
\label{2336mod1}
\end{figure*}
The aim of the mass modelling is the determination of the underlying
density distribution from the observed light profile of NGC 2336.  The
procedure works in several steps: First we determine the light
distribution from the NIR images and decompose it into the three
components disk, bar, and bulge.  By iteration the optimal parameters
of each component like scale lengths, central flux density etc., were
determined by fitting an appropriate model. Since we used NIR images,
this morphological model gives a good representation of the stellar
light and also mass distribution (see Frogel et al., \cite{frogel}). In order
to keep the number of free parameters small we use thickness zero
models for disk and bar, only the bulge model will make use of a 3D profile.

The equation for the flux model of the 3 components is
\begin{equation}
F_{tot}(x,y) = F_{bulge}(x,y) + F_{disk}(x,y) + F_{bar}(x,y).
\end{equation}
For the derivation the mass distribution, one has to take into account
individual mass-to-light ratios $M/L$ for the components:
\begin{eqnarray}
M_{tot} &=& M_{bulge} + M_{disk} + M_{bar}\\
&=& F_{bulge}* (M/L)_{bulge} + F_{disk}*(M/L)_{disk}\\
       &+& F_{bar}*(M/L)_{bar}\nonumber
\end{eqnarray}
For this step we have to consider the dark matter halo component which
has an increasing dynamical influence with increasing radius. However,
for spiral galaxies of medium and high luminosity the dynamical mass
within the optical diameter of the galaxies is strongly dominated by
the disk (e.g. Salucci et al, \cite{salucci}). In the inner zone the dark
matter does not play a dominant role. Since we are mainly interested
in the dynamical influence of the bar, we refrain from modelling the
dark matter component explicitly and include its contribution in
$(M/L)$ of the disk (the bar). This means that our final
$(M/L)_{disk}$ ($(M/L)_{bar}$) should not be directly compared with observed
$(M/L)$ values for spiral disks (bars) since, besides the population effect,
it contains a contribution from the dark matter halo.
This procedure was chosen to include the dark matter in the
corresponding potentials in a simple way, without making further
assumptions about the spatial distribution of the dark matter. 
Therefore, we
hereafter denote the mass-to-light ratios by $C_{bulge}, C_{disk},
C_{bar}$. Equation 3 becomes
\begin{equation}
M_{tot} = F_{bulge}* C_{bulge} + F_{disk}*C_{disk} + F_{bar}*C_{bar}.    
\end{equation}
$C_{bulge}$ is nearly unaffected
by dark matter contributions. Since the morphological model is based on 
J-band images, we set 
\begin{equation}
(M/L)_{J,\,bulge} = C_{bulge} \stackrel{!}{=} 1.
\end{equation}
In a next step we define $C_{disk}$ and $C_{bar}$ as relative
mass-to-light ratios, compared to the bulge.  We start the dynamical
models with a set of start values for $C_{bulge}, C_{disk}, C_{bar}$.
These parameters will be optimized by an iterative comparison between
the kinematical models and the spectroscopic observations, see section
9.  The calibration of the total mass will be performed via the
observed circular velocities in the outer parts of the disk, see
section 10.

\subsection{Profile types}
The following types of luminosity profiles are used for the
luminosity decomposition: 

\subsubsection{Disk}
For the disk, an {\em exponential} profile with the surface
density distribution
\begin{equation}
\Sigma_{ED}=\Sigma_0\cdot
e^{-\frac{r}{r_d}}
\end{equation}
is used with $\Sigma_0$ as the countrate in the center and the scale
length $r_d$ determining the slope of the luminosity profile.  Due to
its inclination the disk appears elliptically projected with an
apparent axis ratio $b/a_d$ and a position angle $PA_d$ of its
apparent major axis.  Free parameters therefore are:
\begin{itemize}
\item
the central intensity $\Sigma_0$ 
\item
the scale length $r_d$
\item
the apparent axis ratio $b/a_{d}$
\item
the position angle $PA_{d}$
\end{itemize}
The apparent axis ratio $b/a_{d}$ (i.e., the inclination $i$) defines --
together with $PA_{d}$ -- the $LON$ for the disk that will be used as
deprojection axis for all three components.

\subsubsection{Bulge}
The central part of NGC\,2336 is modelled by a profile
of generalized
Hernquist type (see Dehnen~\cite{dehnen}) with the density profile
\begin{equation}
\rho_q(m)=\frac{(3-\gamma)M}{4\pi
q}\frac{r_b}{m^\gamma}\frac{1}{(m+r_b)^{4-\gamma}}
\end{equation}
with
\begin{equation}
m^2=x^2+y^2+\frac{z^2}{q^2}\quad .
\end{equation}
$q=0$ yields the flat disk case, $q=1$ results in a spherical
distribution.

With $r_b$ as a scale length and $M$ as the integrated mass, the
central slopes of the densities are $\propto r^{-\gamma}$.
This family
of profiles include several special cases: The steep Jaffe (\cite{jaffe}) type
($\gamma=2$), and the Hernquist (\cite{hernquist}) type
($\gamma=1.0$). Moreover, this profile type is a good approximation for the classical 
de Vaucouleurs (\cite{devaucouleur}) profile
($\gamma=1.5$).
Free fit parameters are:
\begin{itemize}
\item
the scale length $r_b$
\item 
the total mass $M$
\item
the profile parameter $\gamma$
\item
the vertical
flattening of the bulge $q$
\item
the apparent axis ratio $b/a_b$
\end{itemize}
For the position angle of the bulge, $PA_b$, the same value as for the
disk is used, since in disk galaxies the rotation axes of bulge and
disk generally coincide. Nevertheless, $b/a_d$ and $b/a_b$ may differ
due to the different thickness of the components.

The projected luminosity density of the bulge that can be observed
in the sky is obtained by evaluating the integral
\begin{equation}
\Sigma(r)=2\int_R^{\infty}\frac{\rho(r)r}{\sqrt{r^2-R^2}}dr,
\end{equation}
and is expressible in terms of elementary functions for integer
$\gamma$ only (see Dehnen et al.~\cite{dehnen} for further details). 
\subsubsection{Bar}
A Ferrers (\cite{ferrers}) bar with the surface density profile
\begin{equation}
\Sigma (m^2)
= \left\{ \begin{array}{r@{\quad:\quad}l} \Sigma_0(1-m^2)^n & \hbox{if}\, m^2
<1 \\ 0 & \hbox{if}\, m^2 \ge 1 \\ \end{array} \right.
\end{equation}
with
\begin{equation}
m^2=\left(\frac{x}{a}\right)^2+\left(\frac{y}{b}\right)^2, \quad a>b\ge 0
\end{equation}
is used.
The total mass of the 2D bar obtained by integration is
\begin{equation}
M_{bar}=\pi a b \Sigma_0\frac{\Gamma(n+1)}{\Gamma(n+2)}
\end{equation}
(cf.~Matthias \cite{matthias}).
Free parameters that have to be adjusted during the fitting procedure
are:
\begin{itemize}
\item
the apparent bar major axis $a$
\item
the apparent axis ratio $b/a$
\item
the central surface
density $\Sigma_0$
\item
the exponent $n$ that determines the density
distribution of the bar
\item
the position angle of the bar major axis, $PA_{bar}$
\end{itemize}
In our models, all Ferrers bars will be used with $n=2$
for simplicity reasons. 

\subsection{Results}
\begin{table}[b]
\begin{center}
\begin{tabular}{|l|c|c|c|}\hline
&\bf disk&\bf bulge&\bf bar\\ \hline
scale length [kpc]&$r_d=4.76$&$r_b=1.58$&$a=4.17$\\
scale length [arcsec]&42.9&14.2&37.5\\[0.5ex]
apparent $b/a$&0.63&0.85&$0.5\pm 0.1$\\
$PA$ of major axis &$175^{\circ}$&$175^{\circ}$&$120^{\circ}$\\
relative contribution& 66.0\%&21.0\%&14.0\%\\
profile parameter $\gamma$&&0.8&\\ \hline
\end{tabular}
\end{center}
\caption{\label{2336modtab1}\em Parameter values for the decomposition
model of NGC\,2336. $PA$ denotes the orientation of the major axes of
the three components (counter-clockwise). Due to its intrinsically
rounder form compared to the disk, the bulge exhibits a larger $b/a$
ratio although the major axes of both components are oriented in the
same direction.}
\end{table}
The model parameters are adjusted iteratively in order to minimize the residuals that
remain after the subtraction of the model from the observed luminosity
distribution.  
The model of NGC\,2336 that fits the observed light distribution best is shown in
Fig.~\ref{2336mod1}, together with the remaining residuals. The
parameter values obtained from the fit are given in Table
\ref{2336modtab1}. 
\begin{figure}[h]
\resizebox{\hsize}{!}{\includegraphics*{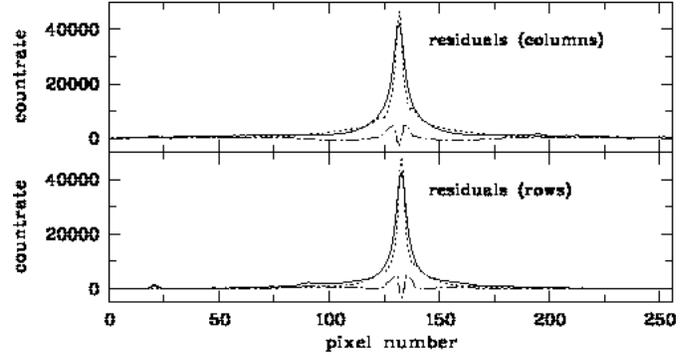}}
\caption{\em
Image slice through the three central pixel columns (upper plot) an
rows (lower plot) of the original $J$-band-image (solid line), the
model (dotted line) and the remaining residuals (dash-dotted
line). For further explanations see text.
}
\label{2336res1}
\end{figure}

As can be seen from the residuals in Fig.~\ref{2336mod1}, modelling
the central regions is difficult: residuals in the bulge area are
still clearly visible after the subtraction of the model.  This is a
consquence of the fairly poor spatial sampling of the bulge area in
the NIR images.  In the same figure we notice that the spiral arm
pattern is still present, because
an average disk was subtracted generating positive residuals in
the spiral arms and negative ones in the interarm regions.

A quantitative impression of the quality of the fit is presented in
Fig.~\ref{2336res1}, where the three central pixel columns and rows of
every image in Fig.~\ref{2336mod1} are
averaged.
The overall fit to the luminosity distribution of NGC\,2336, though
revealing small differences between the model and the observations, is
sufficiently good.  

The morphological decomposition yields the result that NGC\,2336 is
dominated by the disk component, which contributes $\approx 66\%$ to
the total luminosity, with the remaining light coming roughly equally
from the bulge and the bar component.

\subsection{Deprojection}
Because of the inclined position of NGC\,2336 in the sky, the 2D model
of the surface brightness has to be deprojected in order to obtain the
intrinsic bar axis value, $b/a_{bar,\, intr}$.  We use the
deprojection angles $\phi ,\,\theta ,$ and $\psi$, according to the
definition given by Goldstein(\cite{goldstein}), with $\theta$ being
equivalent to the inclination $i$, and $\psi$ representing the offset
between $a_{bar,\, app}$ and the $LON$.
$\phi$ is only necessary if a
non-axisymmetric component (a bar) is present and is used to give the
bar the desired orientation before deprojecting the disk.
The results of the deprojection are shown in
Table \ref{depro1}.
\begin{table}[h]
\begin{center}
\begin{tabular}{|l|c|}\hline
$l.o.n.$&$180^{\circ}$\\
$i=\theta$&$59^{\circ}$\\
$b/a_{bar,\, app}$&$0.5$\\
$b/a_{bar,\, intr}$&$0.28$\\
$\phi$&$80^{\circ}$\\
$\psi$&$104^{\circ}$\\ \hline
\end{tabular}
\end{center}
\caption{\label{depro1}\em Deprojection parameters for
NGC\,2336. The Euler angle $\theta$ is equal to the observed
inclination of the galaxy, $i$, while the Eulerian $\psi$ denotes the
bar offset from the $LON$.}
\end{table}
The large difference between the apparent and the
intrinsic bar axis ratio in Table \ref{depro1} originates from the
orientation of the bar in NGC\,2336 with respect to the $LON$:
$a_{bar}$ is nearly orthogonal to
the $LON$ around which the galaxy plane is rotated during the
deprojection, so the minor axis of the bar remains unchanged. The
enlargement of $a_{bar}$ then leads to a decrease of $b/a_{bar,\,
intr}$.

Since the deprojection angles are three additional free parameters,
the models will have to be tested for possible changes in the
predicted kinematics that are caused by errors of $\phi$, $\theta$,
and $\psi$.  The most sensitive parameter is probably $\psi$: It does
not only include a possible misalignment of the $LON$, but also a
wrong determination of the bar offset from this line. In case of
NGC\,2336, the latter is important due to the orientation change of
the bar isophotes towards larger radii (isophotal twist). To keep
the number of free model parameters as small as possible, $\phi$ and
$\theta$ will remain fixed for all models with their photometrically
derived values being used. $\psi$ will be varied independently, the
results are presented in Section 10, where the effects of parameter
changes are discussed in detail.

A last correction has to be made for the projection effects of the
flattened spherical bulge of NGC\,2336. An infinitesimally thin disk
that is viewed pole-on represents an intensity $I$ that changes to
$I_{\theta}=I/\cos\theta$ when viewed under the inclination
$\theta$. In contrast, a spherical bulge would look the same from
every direction. It is therefore possible that a moderately flattened
bulge introduces further uncertainties in the conversion factor
$C_{bulge}$ defined in Section 3 which converts countrates into
luminosities.  These uncertainties would be caused by changed
inclination corrections for a non-spherical bulge.  It will be
discussed in section 10 that the exact shape of the bulge does not
affect the kinematics of the NGC\,2336-models significantly, since for
reasonable flattening ratios $q=0.4\pm 0.2$ the projected mass density
in the disk plane exhibits only minor changes.

\section{Kinematical Observations}
Spectroscopic observations were performed to determine the velocity
fields of stars and gas in NGC\,2336. By means of longslit
spectroscopy of high spatial and moderate spectral resolution the
radial velocities of stars and gas clouds were measured up to the
outer regions of the disk ($r\ge 100''$) with high $S/N$-ratios.

\noindent
{\em Emission-line kinematics}:\\
For the kinematics of the warm HII gas component we used the light
from the $[NII]$- and $H_{\alpha}$-emission lines in the red
wavelength range.  Since NGC\,2336 is an intermediate-type galaxy, the
disk contains enough gas for emission-line-spectroscopy even in the
outer parts. Radial velocities were obtained by cross-correlating
single rows of the longslit spectrum with the extracted central line
from the core region of NGC\,2336. The velocity dispersion of the warm
gas lies far below the instrumental resolution and was therefore not
studied.

\noindent
{\em Absorption-line kinematics}:\\
The spatial distribution of the stellar velocities and velocity
dispersions was examined by using absorption line spectra in the green
wavelength range ($4500-5500$\AA). Using strong absorption lines like
MgI at $\lambda\lambda_{rest}$=5167\AA, 5173\AA, 5184\AA, CaI lines
at $\lambda\lambda_{rest}$=5262\AA, 5270\AA,  and several Fe lines, the
longslit spectra of NGC\,2336 were cross-correlated with the spectrum
of suitable template stars (e.g.~of type $G8III$, $K0III$, $K3III$, $K5III$). The stellar velocity
dispersions are much higher than those of the warm gas, they exceed
the spectral resolution of our instrument configuration and can be
studied in greater detail, at least in the central regions.

All spectra were obtained during an observing run in March 1997 using
the MPIA $3.5$m-telescope at Calar Alto (Spain) with the
TWIN-spectrograph. Due to a beamsplitting mirror with a crossover
wavelength of $\lambda_{cr}=5500$\AA, this device was capable of
observing the blue and the red channel simultaneously. The cameras in
both channels were equipped with $2048\times 1024$ SITE-CCDs with a
pixel size of $15\mu$m. With a scale length of $1''=178\mu$m along the
slit and the correction for the ratio
$f_{coll}/f_{cam}=6.34$, we obtained a spatial resolution of
$0.56''$/pixel. Gratings with a dispersion of $36$\AA/mm were used,
the grating angles were adjusted such that the central wavelengths
became $\lambda_{c,\,blue}=5100$\AA~in the blue channel and
$\lambda_{c,\,red}=6600$\AA~in the red channel.

To achieve a $S/N$ ratio as large as possible, the slit width was
adjusted to $3.4''$. With our instrument setup and grating angles, the
corresponding slit image was $4.5$ pixels in the blue channel and
$4.05$ pixels in the red one.

NGC\,2336 was observed in 4 slit positions along the major and minor
axes of disk and bar ($a_{disk},\, b_{disk},\, a_{bar},\, b_{bar}$).
The corresponding position angles were
$PA=5^{\circ},\,94^{\circ},\,28^{\circ},\,118^{\circ}$.
Exposure times were the same for emission and absorption spectroscopy:
$2\times 3600$sec in each slit position.  The data reduction process
for both, the emission and the absorption line spectra, used a
standard pre-reduction procedure containing the subtraction of the
bias, the time-dependent linear dark current, and a flatfield division.
The correction for the light distribution along the slit (including
vignetting effects of the camera optics) was made by means of a
slit profile obtained from a series of skyflat exposures.

The subsequent main reduction process of the stellar spectra
consisted of the following steps:
Removal of saturated pixels (cosmics), wavelength calibration
(equidistant in $v_r\propto
\Delta\lambda/\lambda$) via a rebin in $\log\lambda$, sky subtraction,
continuum subtraction, etc.~.
In a last step, these spectra were compared with
correspondingly prepared spectra of the template stars. We used a
hybrid method of Bender (\cite{bender}) which combines
cross-correlation and Fourier quotient evaluation . This method is
less sensitive to template mismatching and does not necessarily assume
Gaussian broadening functions.  From this procedure, the stellar radial
velocity curves were obtained. Velocity dispersions were computed
using the broadening of the maximum of the correlation function.

The data reduction of the emission line spectra was roughly the same,
except for the fact that no template star, but the central line of the
galactic spectrum was used for correlation, and for the slightly
different correlation program being used. 

\begin{figure}[h]
\resizebox{\hsize}{!}{\includegraphics*{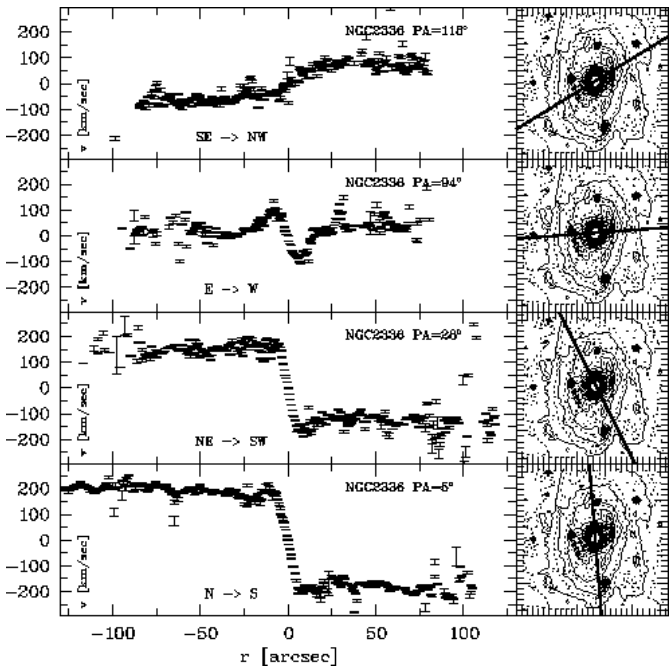}}
\caption{\em Gas rotation curves of NGC\,2336 with a spatial resolution of 
$0.56''$/pixel. For each of the 4 slit orientations 
the position angles ($PA$) are given.
The corresponding slit positions in the sky are indicated by the
labels below the rotation curves and by the thick line in the
isophotal plots ($J$-image). Due to the limited slit length
there is a cutoff e.g.~at $PA=5^{\circ}$ at radii $\ge +105''$.
Notice that the $S/N$-ratio is remarkably high, even in the 
outer regions of the disk (at $100''$).} \label{2336rote}
\end{figure}

\begin{figure}[h]
\resizebox{\hsize}{!}{\includegraphics*{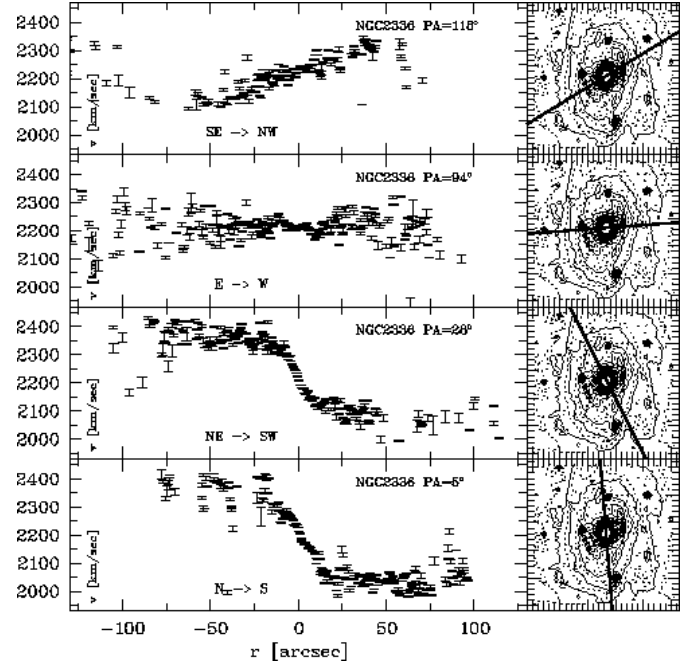}}
\caption{\em Stellar rotation curves of NGC\,2336. Legends are the same 
as in Fig.~\ref{2336rote}. The $S/N$-ratios of the stellar rotation curves 
are lower than those of the gas rotation curves. The 
general kinematic behaviour of the stars differs significantly from 
that of the gas (for further details see text).} \label{2336rota}
\end{figure}

\begin{figure}[h]
\resizebox{\hsize}{!}{\includegraphics*{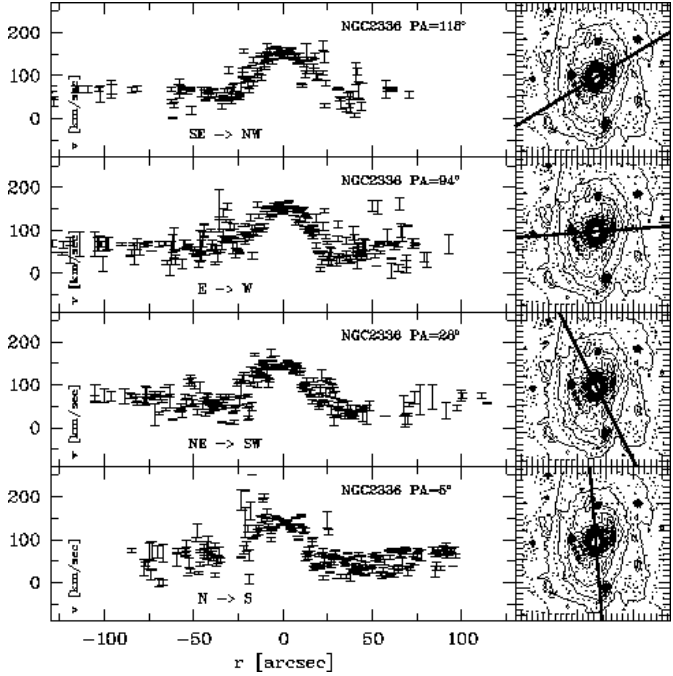}}
\caption{\em Stellar velocity dispersions of NGC\,2336. Legends are
the same as in Fig.~\ref{2336rote}. Only the 
velocity dispersions of the bulge reach values well above the 
instrumental resolution. The disk cannot be detected.}
\label{2336dispa}
\end{figure}

To collect as much light as possible, we chose a slit width of $3.4''$
for all observations which of course affects the spectral resolution:
With $0.54$\AA/pixel (blue channel), the nominal resolution for
measuring velocity dispersions was $\sigma=31.7$km/sec/pixel.
Projected onto 4.5 pixels (blue channel), the slit width became
$\sigma=4.5$pixel$\times 31.7$km/sec/pixel $\approx 140$km/sec in
velocity space.  But since we obtained the velocity dispersions from
measuring the broadening of the maximum of the correlation function
only, the achievable resolutions are much better: A lower threshold of
$\sigma_{eff}\approx 65$km/sec obtained for our stellar spectra is
only marginally worse than the Nyquist-limited spectral resolution of
slit.  The resulting rotation curves of gas and stars and the stellar
velocity dispersion curves are shown in Fig.~\ref{2336rote},
\ref{2336rota} and \ref{2336dispa}.


Considering the gas rotation curve for a slit orientation near the
kinematical $LON$ (Fig.~\ref{2336rote}, $PA=5^{\circ}$), we notice the
steep rise of  the radial velocities to values of $|v_{rot}|\approx
185$km/sec within $13''$ only, corresponding to $1.45$kpc. This ascent
in the inner part of NGC\,2336 is caused by a centrally condensed and
moderately massive bulge component. The presence of the bar is
revealed by small humps (overshootings) in
the rotation curve at a distance of $r\approx 13''$ from the center,
after which the velocities decline again
($|\Delta v_{rot}|\approx -45$km/sec).
At distances from the center greater than $33''$,
$v_{rot}$ is constant, indicating that NGC\,2336 does not contain a
disk component with a large scale length that would cause a
considerable increase of $v_{rot}$ beyond the regions dominated by the
bar.

In contrast to that, the stellar rotation curves for the same slit 
orientation ($PA=5^{\circ}$) in Fig.~\ref{2336rota} show a different 
behaviour: The final values of $v_{rot}$ are not reached within 
of $r\approx 24''$ (2.7kpc), although the final values of $v_{rot}$
are indeed the same. Therefore the central slopes of these two gas rotation
curves in the innermost regions are significantly shallower than those
of the stellar rotation curves.

This behaviour originates from the qualititatively different orbits
which stars and HII-clouds are moving on: Because of the non-neglectible
cross-section of the spatially extended HII-regions, they accumulate
preferentially on non-intersecting orbits in the disk.  If
HII-regions would populate self-intersecting orbits with loops,
collisions would remove them from these paths after only a few
rotation periods of the bar. As a result, intersecting orbits do not
play any role for the kinematics of the warm gas component. In other
words, the radial component of the motions of the HII-regions will be
dissipated with time so that only circular or nearly circular orbits
survive. This means that -- viewed from the corotating reference frame
of the bar -- in our stationary models the HII-clouds will populate
closed non-intersecting orbits of elliptical or circular shape.

In contrast to that, stars can be regarded to as point-like test
particles that do not collide on timescales of $H_0^{-1}$. In general,
stars will therefore also populate orbits with a high contribution
of radial motion. This is especially true for the stellar bulge which
-- like an elliptical galaxy -- mainly consists of disordered motion.
Therefore the integration over many stellar orbits along the line of
sight results in a reduced radial velocity and high velocity
dispersion values in the bulge region (cf.~Fig. \ref{2336dispa}).
This causes the different
slopes of the stellar and HII-rotation curves in the inner regions of
NGC\,2336 in Fig.~\ref{2336rote} and \ref{2336rota}.

An even larger difference between the motion of stars and gas occurs
in case of a slit position along the minor axis of the disk
($PA=94^{\circ}$) in Fig.~\ref{2336rote} and \ref{2336rota}. The
humps and dips in the gas rotation curves with $\vert \Delta
v_{rot}\vert = 110$km/sec at a distance of $r=7''$ (0.6kpc) from the
center indicate high streaming velocities along the bar.
Those humps and dips are
also present in the stellar rotation curve in Fig.~\ref{2336rota},
which is a necessary condition for the existence of the bar structure,
since stars constitute the bar. But as is clearly visible, those
overshootings are much smaller, since the innermost regions of
NGC\,2336 are dominated by the highly disordered radial motions of the bulge
stars which nearly cancel out during the integration along the line of
sight.  To summarize, HII-clouds can been regarded to as ideal test
particles to trace the underlying gravitational potential of the
bar. The latter is mainly generated by the stars, carrying most of
of the total visible mass in normal intermediate-type disk galaxies.

The comparison between radial velocities of gas and stars for
$PA=118^{\circ}$ (along the major axis of the bar) in
Fig.~\ref{2336rote} and \ref{2336rota} yields no qualitative
difference, except for the $S/N$-ratio being again low for the stellar
kinematics, which prevents us from comparing the outer regions with
$r\ge 50''$. Both curves (gas and stars) show similar solid body
rotation over the whole length of the bar.

As shown in Fig.~\ref{2336dispa}, NGC\,2336 reaches a central velocity
dispersion of $\sigma = 160$km/sec. In all velocity dispersion curves
the bulge extends up to radii of $r\approx 40''$. This means that the
apparent bulge axis ratio cannot be much smaller than 1, with
$b/a\approx 0.8$ as a lower boundary, taking the velocity dispersion
curves at $PA=5^{\circ}$ (disk major axis) and $PA=94^{\circ}$ (disk
minor axis) as reference. It is not possible to determine this axis
ratio here more exactly, but a kinematically derived condition $b/a\ge
0.8$ agrees well with the photometrically derived value of $b/a=0.85$.

\section{Potentials}
To derive the common potential of disk, bulge, and bar, we make the
following approximation: Disk and bar are collapsed to their plane of
symmetry, their potential is calculated in 2D. The bulge is considered
as an oblate spheroid, but the potential of this 3D mass distribution
is computed only in the disk plane.  The following section deals with
the derivation of the 2D potentials from the morphologically derived
luminosity profile and their implementation into the FORTRAN-/C-orbit
integrator.

In general, potentials and density distributions are connected by
Poisson's equation $\Delta\Phi(\boldmath{x})=4\pi G\rho(\boldmath{x})$.
For the Ferrers bar, the exponential disk and the generalized
Hernquist bulge, no analytical closed expressions for their 2D
potentials are available
(except for some special cases),
so they have to be evaluated by means of
numerical representations or expansion series.

\subsection{Disk}
It is not appropriate for the potential calculation to consider the
disk as a collapsed 3D sphere with thickness zero, since one
encounters double integrals that cannot be solved easily. As a better
approach, we use Bessel functions to derive the disk potential
(described by Binney and Tremaine~\cite{binney}, originally developed
by Toomre~\cite{toomre}). With the ansatz
$\Phi_k(R,z)=e^{-k\vert z\vert}J_0(kR)$,
the Laplace equation $\nabla^2\Phi=0$ can be solved except for the
2D plane of the disk, where the gradient is discontinuous. Using the
Gauss theorem, the mass density that generates the gradient
discontinuity is found to
$\Sigma_k=-\frac{k}{2\pi G}J_0(kR)$
with $J_0$ as cylindrical Bessel function of 0th order.
With the Hankel transform $S(k)$ defined by
\begin{equation}
\Sigma(k)=\int_0^{\infty}S(k)\Sigma_k(R)dk,
\end{equation}
the potential is given by
\begin{equation}
\Phi(R,z)=\int_0^{\infty}S(k)\Phi_k(R,z)dk.
\end{equation}
It can be shown that for the exponential disk with\\
$\Sigma(R)=\Sigma_0e^{-R/R_d},$
$S(k)$ takes the form
\begin{equation}
S(k)=-\frac{2\pi
G\Sigma_0R_d^2}{[1+(kR_d)^2]^{\frac{3}{2}}}\quad .
\end{equation}
For the 2D disk ($z=0$) and with $y\equiv R/2R_d $, this results in
$\Phi(R,0)=-\pi
G\Sigma_0R[I_0(y)K_1(y)-I_1(y)K_0(y)]$.
$I_n$ and $K_n$ are modified Bessel functions of 1st and 2nd
kind. This expression is implemented in the program used for the
calculation of closed particle orbits.

\subsection{Bulge}
For the generalized Hernquist bulge luminosity profile, there are no
closed analytic expressions for the corresponding $\Phi(r)$ (see
Dehnen~\cite{dehnen}). Therefore, our orbit integrator includes
FORTRAN-routines written by Walter Dehnen (Oxford) which allow a
numerical treatment.  The 3D bulge potential for the generalized
Hernquist profile takes the form
\begin{equation}
\Phi_q(R,z)=-\frac{GM}{2r_b}\int_0^{\infty}\frac{\tilde\psi(\tilde
m)d\tau}{(\tau+1) \sqrt{\tau +q^2}}
\end{equation}
with $q$ as the flattening parameter,
\begin{equation}
\tilde m=\sqrt{\frac{R^2}{\tau+1}+\frac{z^2}{t+q2}}
\end{equation}
and
\begin{equation}
\tilde \psi(\tilde m)=\frac{1}{2-\gamma}\left[1-\frac{\tilde
m^{3-\gamma}+(3-\gamma)\tilde m^{2-\gamma}}{(\tilde
m+1)^{3-\gamma}}\right]\quad .
\end{equation}
The bulge is the only component with a 3D potential implemented into
the numerical orbit integrator programme (disk and bar are considered with
the corresponding 2D expressions). Since the flattening parameter $q$
determines the mass density projected onto the 2D plane,
$\Phi_{bulge}$ will be sensitive to changes of that parameter.  

\subsection{Bar}
To illustrate the derivation of the bar potential $\Phi_{bar}$, we
first derive a 3D expression for the potential of a general oblate
spheroid and reduce that expression to the 2D case afterwards. The
resulting 2D potential allows a very simple numerical treatment of the
bar. 

A generalized 3D-disk can be expressed by means of an expansion method
described by Binney and Tremaine (\cite{binney}) that uses oblate
spheroidal coordinates. Those coordinates use the
$\phi$ from cylindrical coordinates, but replace $R$ and $z$ by
$R = \Xi\cdot \cosh u \sin v$, and
$z = \Xi\cdot \sinh u\cos v$, respectively,
with $\Xi$ as a constant parameter determining the range of the
coordinate system. 
The potential of a 3D disk outside the inner 2D area $u=0$, where this
coordinate system would be ambiguous (there exist two $v$-values for
each $R$), takes the form
\begin{equation}
\Phi=\sum_{l=0}^{\infty}\sum_{m=-l}^{l} \Phi_{lm}.
\end{equation}
with
\begin{equation}
\Phi_{lm}(u,v,\varphi)=\left[\frac{V_{lm}}{Q_l^m(0)}\right]Q_l^m(i\sinh
u)Y_l^m(v,\varphi).
\end{equation}
$Q_l^m$ are the Legendre polynomials of 2nd order,
$Y_l^m$ are the spherical harmonics. $V_{lm}$ is a constant that contains
\begin{equation}
\Sigma =\sum_{l=0}^{\infty}\sum_{m=-l}^{l}\Sigma_{lm}
\end{equation}
with
\begin{equation}
\Sigma_{lm}=-\left(\frac{2V_{lm}}{\pi^2G\Delta
g_{km}}\right)\frac{Y_l^m(v,\varphi)}{\vert\cos v\vert} 
\end{equation}
as the generating surface densities that account for the
$\Phi_{lm}$. $g_{km}$ and $G$ are constants.  The generating 3D
potentials $\Phi_{lm}$ are divergent for $R\rightarrow \Xi$, but
$\Sigma$ remains finite. This causes numerical problems when one wants
to compute potential values at $R\cong \Xi$. This restriction limits
the use of the expansion series approximation, because we are
prevented from calculating $\Phi$ at large radii (in the vicinity of
$\Xi$). Nevertheless, the method is generally useful for approximating
potentials with sharp boundaries $R(\varphi)$ (like the Ferrers bar).

In the inner 2D disk (defined by $u=0$ in the spheroidal coordinate
system), which is the only interesting region for calculating the
potential of a 2D bar, the coordinate transformations reduce to
$R=\Xi\cdot \sin v$ and $z= 0$
with $0\le v\le \pi/2$.
The collapsed 2D potentials are no longer $u$-dependent:
\begin{equation}
\Phi_{lm}=V_{lm}Y_{lm}(v,\varphi).
\end{equation}
This simple form is sufficient for our 2D studies and has the
advantage that arbitrary bar forms (e.g.~asymmetric or drop-shaped bars)
can be treated.

According to the initial definition as scale parameter, 
$\Xi$ defines the model unit for the radius,
so $\Xi=1$ is the maximum radius at which the potentials
$\Phi_{lm}$ can be calculated. Scaling of the model according
to different bar lengths or different radii up to which the models
have to be examined is achieved by defining $a_{bar}$ and
$b_{bar}$ as fractions of $\Xi=1$, e.g.~$a=0.7$ and $b=0.3$. Scale lengths of the bulge and the
disk component are then adjusted according to the bar
length. Therefore, in a model with a bar major axis $a$, potentials can
be calculated up to a distance of $1/a$.
For all models a bar length $a_{bar}=0.7$ will be used, the
scalelengths of disk and bulge are then defined according to
$a_{bar}$.
It turns
out that, with the resulting maximum radius of $1/a=1.42$ bar lengths
accessible by the potential expansion series,
the outer Lindblad resonance $OLR$ lies well within the reach of most
of our models.
To restrict the consumption of computational time, we truncate the
expansion series (equation 20) at appropriate values $m_{max}$ and $l_{max}$ that determine
the angular and radial resolution. For most of the models
discussed below we use $l_{max}=50$ and $m_{max}=12$ which ensure
sufficiently resolved potentials.

In a final step,
the total potential of the NGC\,2336-models is obtained by simply
coadding all contributions:
\begin{equation}
\Phi_{total}=\Phi_{disk}+\Phi_{bulge}+\Phi_{bar}.
\end{equation}

\section{Resonances and periodic orbit families}
For small deviations from axisymmetry, barred potentials can be
treated by means of linear theory, in which a star radially oscillates
around its guiding center with an epicyclic frequency $\kappa$. The
guiding center moves around the center of the galaxy with the circular
frequency $\Omega(R)$. In this approximation,
several resonances appear where particles are subject to strong interactions
with the rotating bar potential: the inner Lindblad Resonance ($ILR$),
the Corotation ($CR$), and the outer Lindblad Resonance (hereafter $OLR$, 
for an extended review see Sellwood and
Wilkinson~\cite{sellwood}). However, in case of NGC\,2336 this
approximation is not valid due to the non-neglectible contribution of
the massive bar component to the total potential. But although
single particle orbits can no longer be computed analytically, the
examination of the resonances is still useful when studying the
kinematics.
The locations of the
resonances have to be
determined directly from the analysis of the phase space
(Combes~\cite{combes}).

In all NGC\,2336-models, the typical $x_1$, $x_2$- and $x_4$-families of closed, therefore
periodic orbits exist between the different resonance locii.
The general properties of these orbit families are the same for all
NGC\,2336-models and representative for a disk galaxy with a bar
component contributing roughly 10\% to the total potential, as given
by Contopoulos (\cite{contopoulos}). 

The orbits of the $x_1$-family are elongated along the bar major axis within $CR$, they
are the main orbit family supporting the bar. They cover the range
from the center up to the outer regions of NGC\,2336. Outside $CR$
they become rounder and change their
rotation sense 
from prograde to retrograde,
since particles on such orbits
are moving slower than the bar.
Epicyclic orbits with loops, belonging to higher (n:m)-resonances of
the $x_1$-family, will be populated by stars only and are therefore
excluded from the orbit sets used for the construction
of artificial rotation curves of the HII-gas.
$x_2$-orbits are confined to the region between
the center and $ILR$.
The nearly circular $x_4$-orbits do not support the bar.
Because of their retrograde orientation they are not populated by gas
clouds. They are therefore neglected for the construction of
HII-rotation curves. 

In the rotating reference frame of the bar, the normal energy $E$ is
no longer conserved along the particle trajectories, since virtual
forces appear that add a centrifugal and a Coriolis term to the
equation of motion. Instead of
$E=\frac{1}{2}{\dot {\mathbf r}_{in}^2+\Phi}$
(with $\dot {\mathbf r}_{in}$ denoting the velocity in an
inertial coordinate system, cf.~BT87), the so-called Jacobi energy
$E_J$ is conserved:
\begin{equation}
\frac{dE_J}{dt}=0,
\quad\hbox{
with}\quad
E_J=\frac{1}{2}{\dot {\mathbf r}}^2+\Phi -\frac{1}{2}\vert {{\mathbf \Omega}_p}\times
{\mathbf r}\vert^2.
\end{equation}
$E_J$ will therefore be used as model parameter instead of the normal
energy $E$. 

\section{Model Units}
\subsection{Length Scale}
As described in Section 5.3, model length scales are coupled to the
numerical expansion radius $\Xi=1$ by choosing the maximum radius
the kinematics of orbits shall be computed for. It turns out that for
reasonable mass distributions, $0.7\Xi$ is a good choice for
$a_{bar}$, since the $OLR$ lies then well within the reach of our
models. The length units $LU$ are calibrated using the $NIR$-images.
As given in Section 3.2, the best photometric decomposition model of
NGC\,2336 has a minor axis length of the bar of $b=27.0$ pixels. With
the scale of $0.67''$/pixel of the MAGIC camera, we get
$b=18.09''$. With a $b/a_{bar,\,intr}=0.28$ the length of $a$ becomes
\[
a=18.09''/0.28=64.6''\stackrel{!}{=}0.7\Xi.
\]
With a distance of $d=22.9$Mpc (based on $H_0=100$km/sec/Mpc),
we get
$1''\equiv 111pc$,
which results in the model scale ($LU$)
\begin{equation}
1LU\equiv 92.3''=10.25kpc,
\end{equation}
so the photometrically derived bar of NGC\,2336 has the dimensions
$a=7.2$kpc and $b=2.0kpc$.

\subsection{Mass Scale}
We calibrate the mass scale via the equation for the circular velocity
\begin{equation}
v_c^2=r\frac{d\phi}{dr}=\frac{GM(r)}{r}.
\end{equation}
We use those velocity values that are found far out in the disk
($r\approx 9$kpc) to obtain
\begin{eqnarray}
M_{tot}&=&M_{bulge}+M_{disk}+M_{bar}=\frac{v_{circ}^2\cdot r}{G}\\
&=&F_{bulge}\cdot C_{bulge}+F_{disk}\cdot C_{disk}+F_{bar}\cdot C_{bar}.
\end{eqnarray}
Since the relative mass-to-light ratios $C_{disk}$ and $C_{bar}$ have
to be found by optimizing the kinematical model, this is an iterative
process.  The kinematical effects of varying $C_{disk}$ and $C_{bar}$
are studied by corresponding model sequences, similar to those for the
scale lengths, the pattern speed of the bar, etc.  With this method, the mass
unit $MU$
is determined to
\begin{equation}
1MU\equiv 3.234\cdot 10^9M_{\sun}.
\end{equation}
\subsection{Time scale}
The time scale is adjusted such that $G=1$. This yields
\begin{equation}
1TU\equiv 2.71\cdot 10^8yrs.
\end{equation}

\section{Phase Space Analysis and the Integration of Closed Orbits}
\subsection{The {\em Poincar\'e}-Surfaces of Section}
Stars and gas particles in real barred galaxies are normally not
likely to strictly follow periodic orbits (so their occupation number
should be low), but it can be shown that most of the non-periodic
orbits in barred galaxies are trapped to oscillate about a parent
periodic one (for an extended review on phase space analysis see
Sellwood and Wilkinson \cite{sellwood}). Such orbits are often
referred to as quasi-periodic.

Therefore the examination of periodic orbits is an important tool,
because the structure of the bar is largely shaped by those parent
orbits. To derive the spatial extent of the regions supported by the
parent orbits, the division of phase space has to be studied by means
of numerical methods. For those purposes, cuts through phase space at
a certain orbit energy -- the Poincar\'e-Surfaces of Section
(hereafter $SOS$) -- are widely used.
Positions and velocities are
measured every time a particle crosses the plane $x=0,\, \dot x>0$ (or
$y=0,\,\dot y>0$), thereby constituting the $y,\,\dot y$- (or
($x,\,\dot x$))-Surface of Section at a given Jacobian energy
$E_J$.

\begin{figure*}[t]
\resizebox{\hsize}{!}{
\includegraphics*{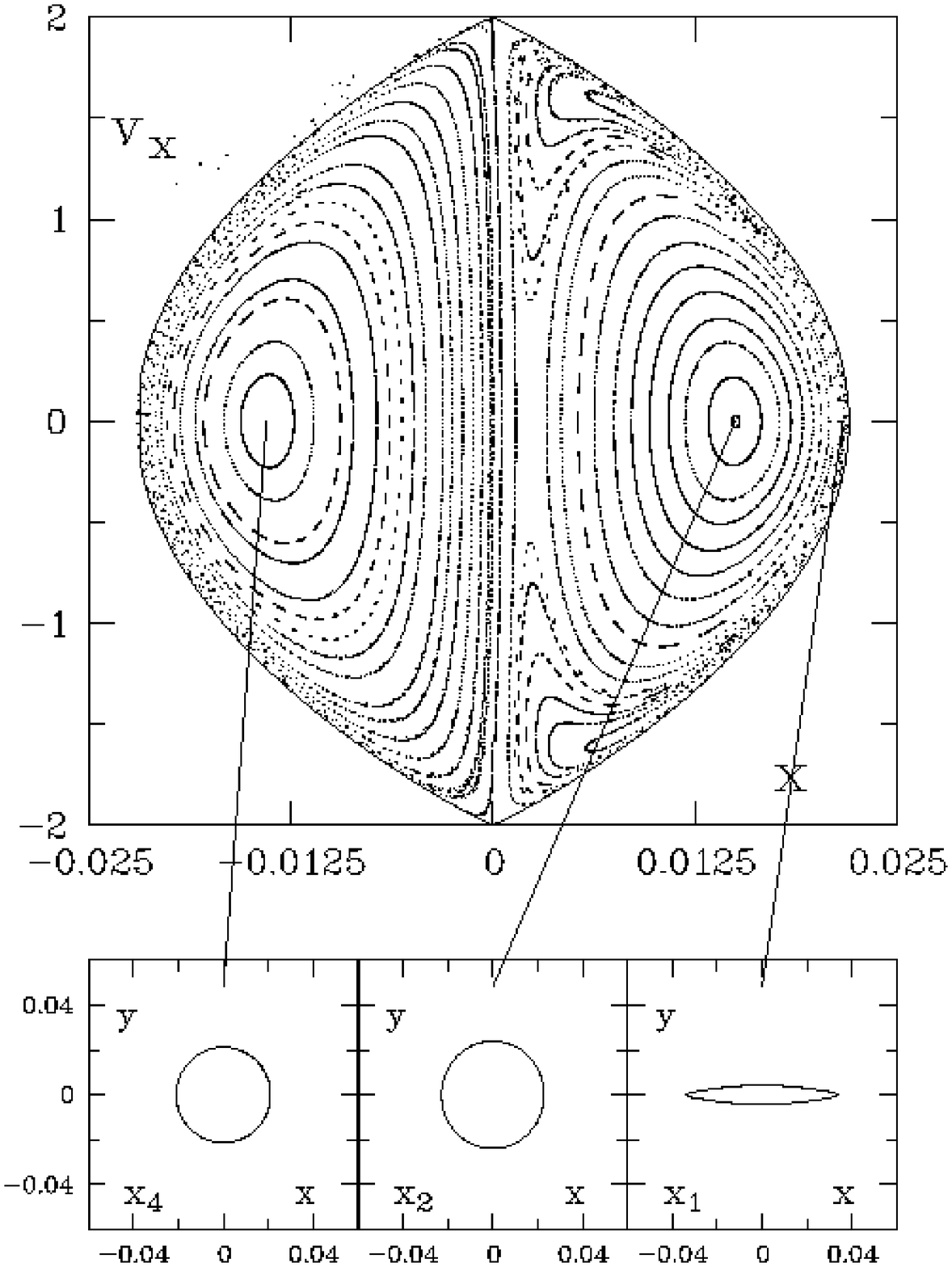}\hspace{0.1cm}
\includegraphics*{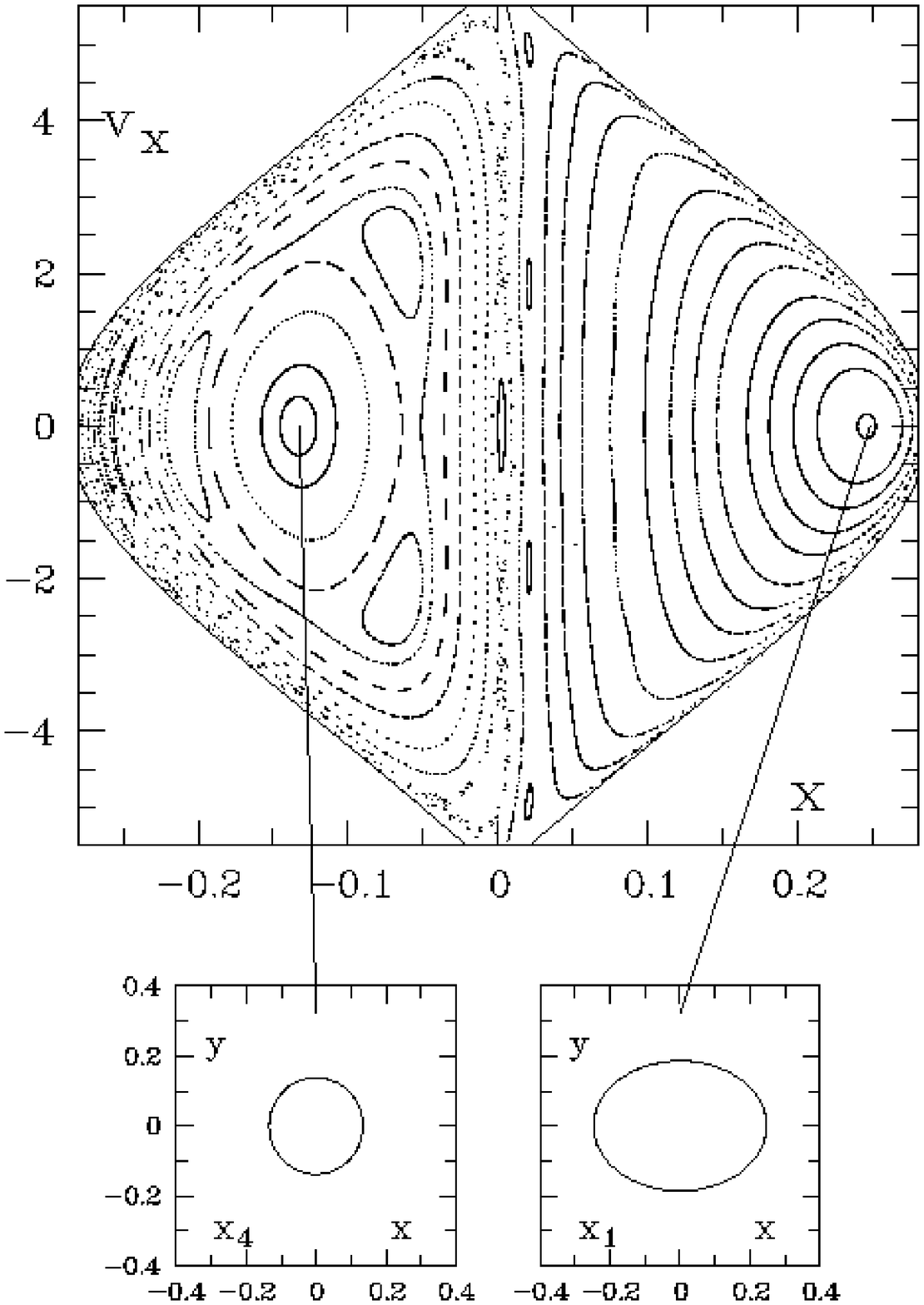}\hspace{0.1cm}\includegraphics*{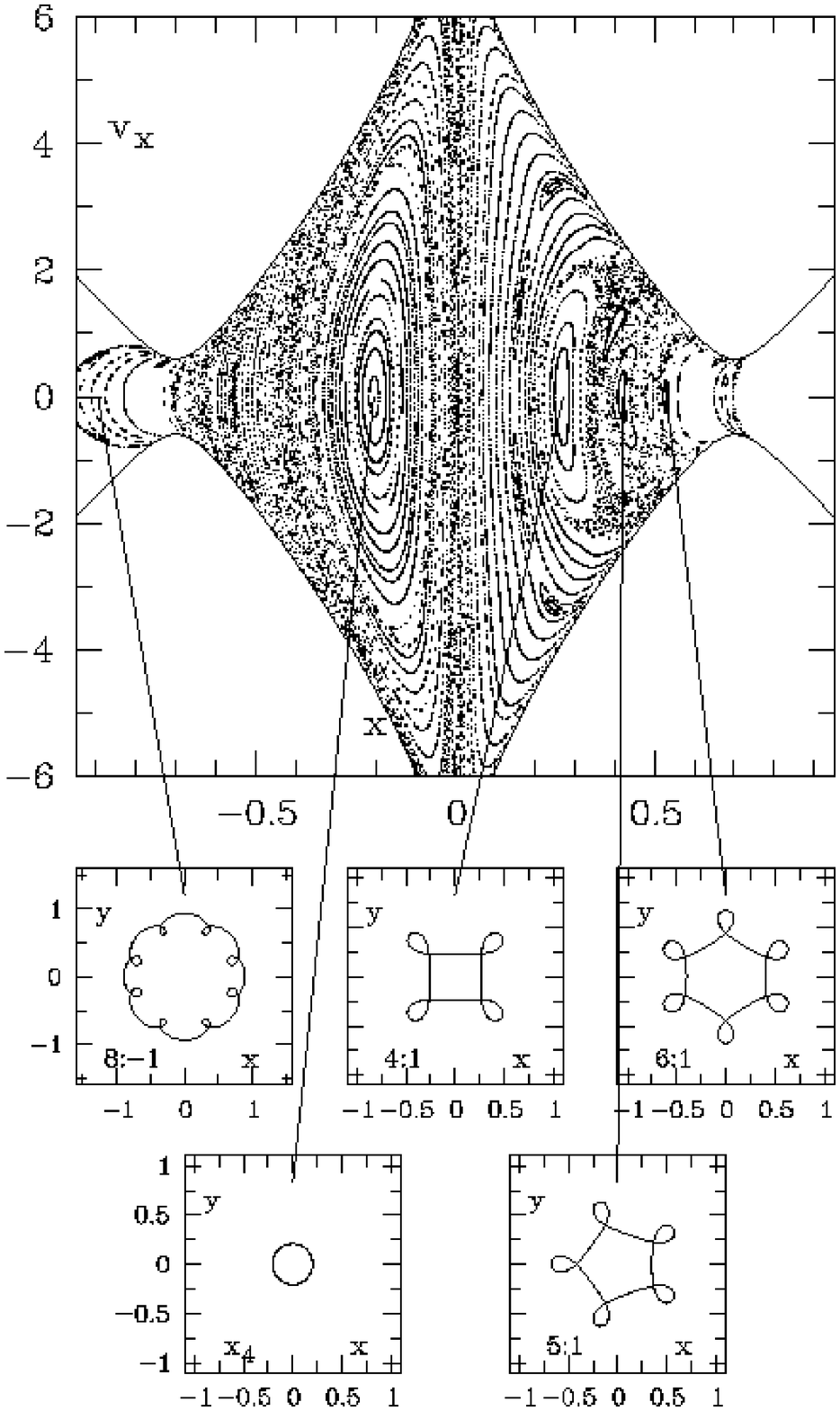}}
\caption{\em
$(x,\dot x)$-Surfaces of Section for a typical NGC\,2336-model with
a moderately flattened ($q=0.6$) bulge. The Surfaces of Section are
computed for Jacobi energies of $E_J=-45.0$ (left), $E_J=-30.0$
(center) and $E_J=-26.3$ (right).  Below every $SOS$, the periodic
orbits of the main families being present in the $SOS$s are
plotted in the $(x,y)$-plane, the solid lines indicating the
positions those orbits can be found at in the $SOS$s above.
}
\label{sos1}
\end{figure*}

The division of phase space in the case of the NGC\,2336-models
exhibits characteristic features which are explained using results of
a model with a moderately flattened ($q=0.6$) bulge (model
parameters are listed in Table \ref{modmb1}).
\begin{table}[h]
\begin{center}
\begin{tabular}{|l|c|}\hline
\multicolumn{2}{|c|}{\bf bar}\\[0.3ex]\hline
$a_{bar}$ [LE/kpc]&$0.7/7.17$\\
$b_{bar}$ [LE/kpc]&$0.21/2.15$\\
$b/a$&$0.3$\\
$C_{bar}$& 1.5\\[1.0ex]
\hline
\multicolumn{2}{|c|}{\bf bulge}\\[0.3ex]\hline
$\gamma$&$0.9$\\
$r_b$ [LE/kpc]&$0.20/2.04$\\
$q$&$0.6$\\
$C_{bulge}$& 1.0\\[1.0ex]
\hline
\multicolumn{2}{|c|}{\bf disk}\\[0.3ex]\hline
$r_d$ [LE/kpc]&$0.46/4.76$\\
$C_{disk}$& 2.5\\[1.0ex]
\hline
$\Omega_p$ [km/sec/kpc]&$17.85$\\ \hline
\end{tabular}
\end{center}
\caption{\label{modmb1}\em
Parameters for a NGC\,2336-model with a moderately flattened bulge and low
$C$-values for bar and disk. All scale lengths are given in
model (LU) and physical units (kpc).
Masses are given in solar mass
units ($M_{\odot}$).
The mass calibration refers to a preliminary calibration according to Sect.~3.
With the selected $\Omega_p$-value, the $CR$ is
placed at the end of the bar: $r_{CR}=1.0a_{bar}=0.7LU$. With a bar axis
ratio of $b/a=0.3$, a value slightly larger than that obtained from
deprojection ($b/a=0.28$) is chosen.  $E_J$ is computed according to
its definition in Section 6, using the model units of Section 7.  The
same units are used for all following tables.
}
\end{table}
The $(x,\dot x)$-Surfaces of Sections for different
$E_J$-values are
displayed in Fig.~\ref{sos1}.
For low $E_J$-values (left plot in Fig.~\ref{sos1}), the phase space
of the NGC\,2336 is dominated by $x_1$- and $x_2$-orbits on the
prograde side ($x>0$) and by $x_4$-orbits in the retrograde half
($x<0$) of the Surface of Section.
Each periodic orbit
is surrounded
by closed invariant curves (hereafter $IC$s) of quasi-periodic orbits.
For this energy well below that of the inner Lindblad resonance
($E_{ILR}$), $x_2$-orbits cover a considerable area in phase
space. This behaviour originates from the centrally condensed model
with a high bar mass that produces a strong $ILR$. Notice that
particles at that $E_J$-values can travel a small region within the
bar only, since the $SOS$-boundary extend only up to $|x|\approx
0.05LU$, while the bar axes lengths are $a=0.7LU$ and $b=0.245LU$,
respectively.

At higher $E_J$-values (central plot in Fig.~\ref{sos1}), the
 retrograde side
remains
nearly unchanged, while on the prograde side $x_2$-orbits are now
completely absent since $E_J > E_{ILR}$. The $x_1$-orbits have become
rounder and a small amount of semi-ergodic motion can be traced that
surrounds the $x_4$-island.  These semi-ergodic orbits, indicating
irregular motions, are always present in models with a massive bar
and hint at a self-regulating mechanism for the bar strength:
The more mass is accreted in the center of a barred galaxy, the
steeper the central potential will become and the higher will be the
percentage of semi-ergodic orbits. But with too much semi-ergodic
motion, a bar will not be able to sustain its length and mass on
longterm scales because of a lack of particles on orbits supporting
the bar.  If therefore two NGC\,2336-models with the same artificial
HII-rotation curves contain different amounts of semi-ergodicity at
comparable energies, the model with the smaller amount of semi-ergodic
motion will be chosen for stability reasons.

Slightly above the $CR$ energy $E_{CR}=-26.5$, the Surfaces of Section
look completely different. At $E_J=-26.3$, the regions outside the
bar become energetically accessible to closed orbits, as can be seen
in Fig.~\ref{sos1} (right plot). But the phase space outside $CR$ is
only partially populated with orbits, which is a tendency that becomes
even more evident when proceeding towards even higher energies.  For
this model, $CR$ is placed at $r_{CR}=0.7$ (at the end of the bar) by
adjusting the pattern speed $\Omega_p$. In the prograde half of the
phase space, numerous islands belonging to (n:m)-families of higher resonant
orbits have replaced the $x_1$-orbit and its surrounding $IC$s.
For the HII-kinematics, those higher resonances are of no
importance due to their small occupation number, which is caused by
their self-intersections leading to a depletion after short times.
They contribute only to the {\em stellar} bar in a significant
manner.  The most important of those islands belongs to the
(4:1)-resonance, which is located at $x\approx 0.28$ in Fig.~\ref{sos1}
(right).  Other higher resonant families (e.g.~the (6:1) family) are
less important.

\subsection{Construction of Artificial Rotation Curves}
Our orbit integrator uses
HII-clouds as test particles to trace the underlying gravitational
potential $\Phi_{total}$ of the NGC\,2336-models. Trajectories are
integrated on a grid-based scheme, which means that in each cell of
the phase space HII-clouds with certain initial conditions
(e.g.~energy, velocities) are started and their trajectories are
followed through the phase space.

For each model, complete sets of closed orbits that belong to the
$x_1$- and (if they are present and/or necessary) $x_2$-family are
computed: The orbit integrator uses the phase space coordinates of the
periodic orbits at various $E_J$-values as starting points and
computes a number of orbits (normally 100 of each family type) belonging to
different particle energies.
The program uses a continuous check for the particle
energy during the integration of every orbit to ensure
conservation of the Jacobian energy $E_J$.
The energy interval for which the orbit sets are
computed is adjusted with regard to the different Lindblad Resonances
in the model (e.g.~orbits between $ILR$ and $CR$
or $CR$ and $OLR$).

All resulting orbit sets are then projected to the sky with the same
inclination and orientation as the galaxy. The orbits generate the
kinematical ``backbone'' of our artificial galaxy. A virtual longslit
is projected to the sky as well. Corresponding to the observations, a
constant slit width of $3''$
is chosen for all models discussed
below. The orbit integrator looks for the points where an orbit
crosses the virtual slit and computes $r$ and $v_{rad}$ for every
crossing point. All crossing points together yield the artificial
rotation curves for a given orientation angle of the virtual slit.
The large number ($\ge 100$) of orbits in each complete set ensures a
sufficient spatial resolution of the artificial rotation curves even
in the central regions of the NGC\,2336-model.

In a last step the rotation curves are convolved with a gaussian point
spread function of constant width ($1.5''$) to correct for the
seeing conditions of the original longslit spectra.  By varying the
position angle of the virtual slit, all position angles from the
observations can be traced.

\section{Variations of the Model Parameters}
The model parameters were varied to check for qualitative and
quantitative changes in the synthetic kinematics of the HII gas. We also
tried to answer the question whether a possible 
best-fitting model could be subject to ambiguities that are
caused by the fact that the variation of two independent model
parameters may have the same effect on the artificial rotation
curves. Therefore, the following parameters were varied in a range
that agrees well with the photometrical observations:

\begin{enumerate}
\item
The bar offset from the line of nodes, $\psi$:
Several cases were calculated with this
parameter varied in the range $\psi=102^{\circ}\ldots 106^{\circ}$
which is well in agreement with the $J$-band-observations. Since the
Euler angles $\theta$ and $\phi$ are subject to very small
deprojection uncertainties only, they remain fixed for all models.
\item
Major and minor axis of the bar, $a$ and $b$: Due to the presence of
another dominant component in the center of NGC\,2336 -- the bulge --
errors may occur when evaluating $b/a_{bar}$. In principle, a larger $b/a_{bar}$
could be compensated by a different choice of the bulge scale length, $r_b$.
\item
The bar mass, $M_{bar}$: The exact determination of $M_{bar}$ is
not easy since the bulge component contributes to the luminosity in
the inner regions as well.
The bar mass (equivalent: $C_{bar}$) was
varied at the expense of the bulge and vice versa.
\item
The bulge scale length, $r_b$:
The fit of the bulge scale length,
$r_b$, to the $J$-band luminosity distribution is difficult because of
the small detector area covered by this component. Variations of $r_b$
in the models are therefore necessary.
\item
The flattening of the bulge, $q$: This parameter is important because
it determines the percentage of the total projected bulge mass that is
concentrated in the disk plane. The value of $q$ could in principle be
estimated directly from the photometric data by evaluating the
different apparent axis ratios of the (flattened) bulge and the
(infinitesimally thin) disk, $b/a_{app,\, bulge}$ and
$b/a_{app,\,disk}$.  But as it is the case for $r_b$, the reliability
of the estimated $b/a_{app,\, bulge}$-value suffers from the low
resolution in the 
bulge area.
In our models, $q$ is therefore varied in the
reasonable range $q=0.2\ldots 0.6$, well consistent with the
photometrically derived apparent axis ratio.
\item
The bulge mass, $M_{bulge}$:
The effects of different bulge masses ($C_{bulge}$-values)
strongly mix up with the results of changing $M_{bar}$ ($C_{bar}$),
therefore models have to be checked for this parameter indepenently.
\item
The disk scale length, $r_d$:
The exact determination of $r_d$ is limited by the size of
the $NIR$-detector that partially omits the outer disk regions. Since
no further information about $r_d$ can be obtained, the
photometrically derived value is used for all models.
\item
The disk mass, $M_{disk}$:
Changing $M_{disk}$ ($C_{disk}$) causes significant velocity changes
in the regions far outside the bar. Its contribution to the total mass
varies according to the masses of bulge and bar.
\item
The bar pattern speed, $\Omega_p$:
This parameter does not
only depend on the radial mass distribution of a model, but can also
be varied independently since no direct observational constraint is
available.
To make the models better comparable when
varying the other parameters, each NGC\,2336-model should have $CR$
placed at the same absolute radius, since certain features in phase
space are spatially connected to the $CR$ radius, e.g.~the regions of
semi-ergodic motion or the occurrence of higher resonant families.
To
achieve a constant value for $r_{CR}$, it is necessary to change
$\Omega_p$ each time the radial mass distributions are changed by
variations of scale lengths and masses.
\begin{table}[h]
\begin{center}
\begin{tabular}{|l|c|c|}\hline
\bf Model&\bf 1&\bf 2\\[1.5ex]\hline
\multicolumn{3}{|c|}{bar}\\
\hline
$a_{bar}$&0.7/7.17&0.87/7.17\\
$b_{bar}$&0.245/2.51&0.245/2.51\\
$b/a$&0.35&0.35\\
$C_{bar}$&1.0&5.0\\[1.0ex]\hline
\multicolumn{3}{|c|}{bulge}\\
\hline
$\gamma$&0.9&0.9\\
$r_b$&0.20/2.04&0.20/2.04\\
$q$&0.4&0.4\\
$C_{bulge}$&1.0&1.0\\[1.0ex]
\hline
\multicolumn{3}{|c|}{disk}\\
\hline
$r_d$ &0.46/4.76&0.46/4.76\\
$C_{disk}$&2.5&2.5\\[1.0ex]\hline
$\Omega_p$ &18.2&19.1\\ \hline
\end{tabular}
\end{center}
\caption{\label{modbm1}\em
Model parameters for two models with different bar masses. The scale
lengths are given in model (LU) and absolute units (kpc).
All other parameters are the same as in Table
\ref{modmb1}. To illustrate the qualitative difference between low and
high bar masses, the bar conversion factor is set to an
(observationally not confirmed) value of $C_{bar}=5.0$ in model 2, 
while model 1 uses $C_{bar}=1$, corresponding to the mass directly
obtained from the morphological decomposition procedure.}
\end{table}
\begin{figure*}[t]
\resizebox{\hsize}{!}{\includegraphics*{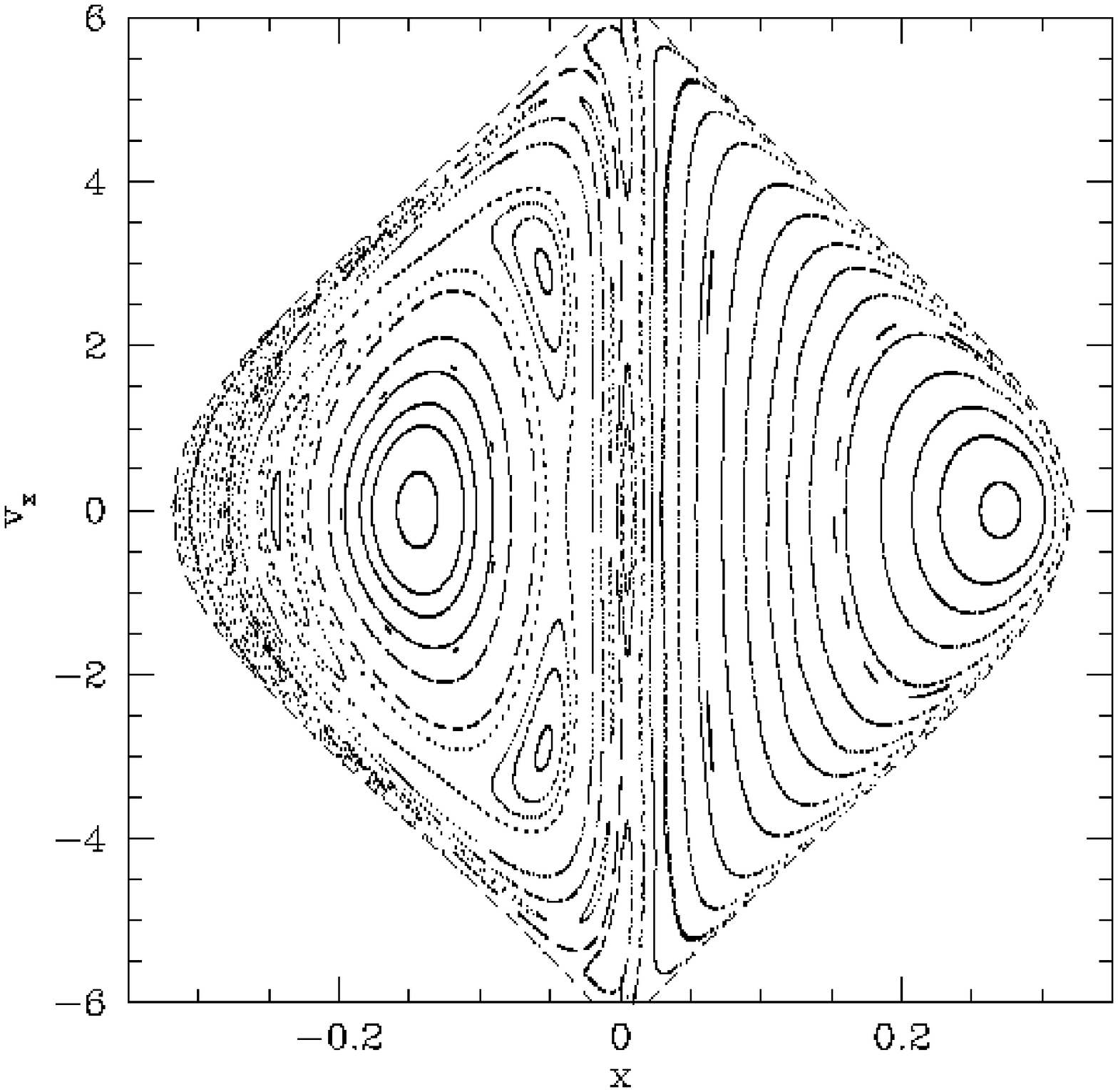}\hspace{0.1cm}
\includegraphics*{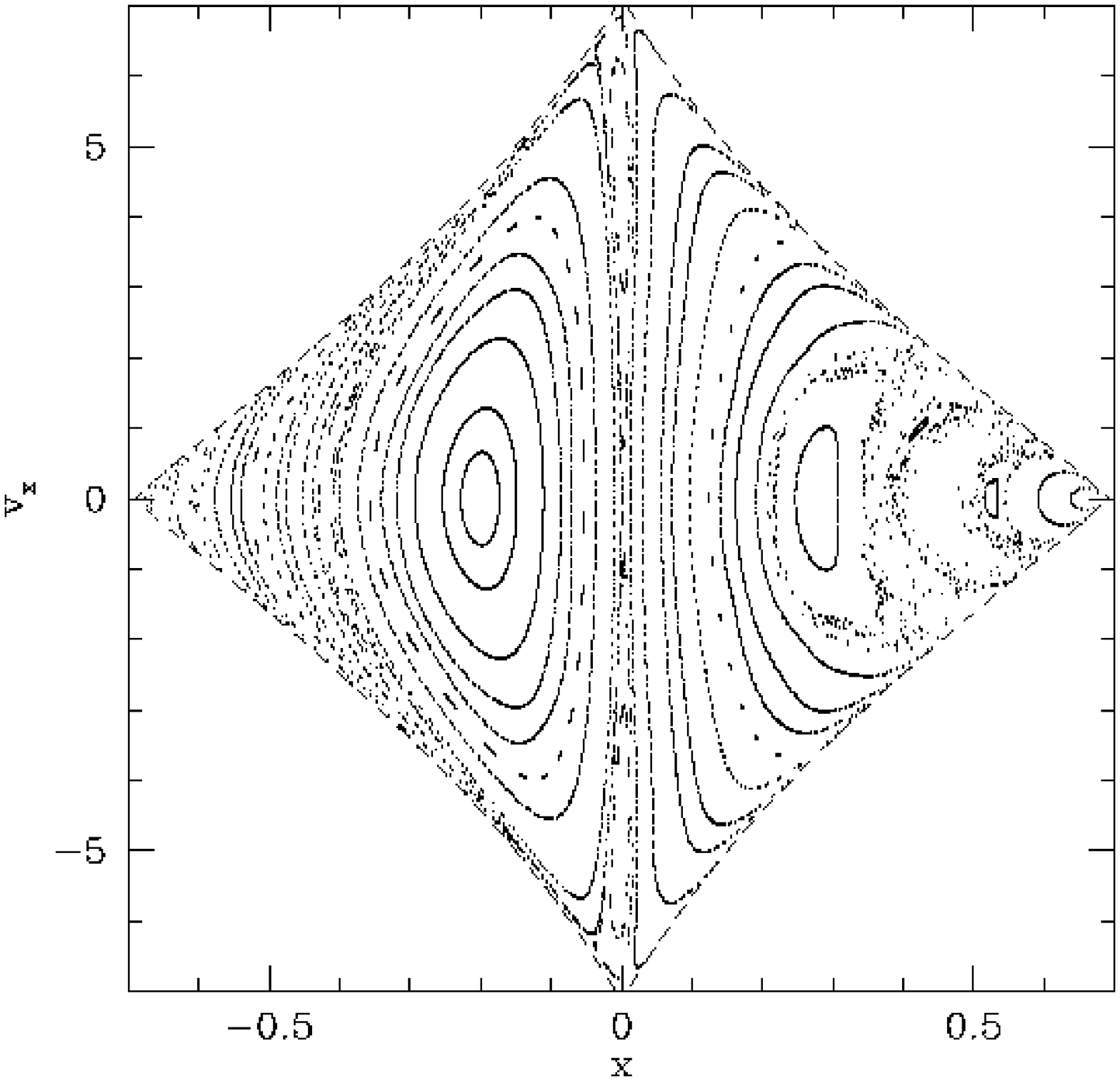}\hspace{0.1cm}\includegraphics*{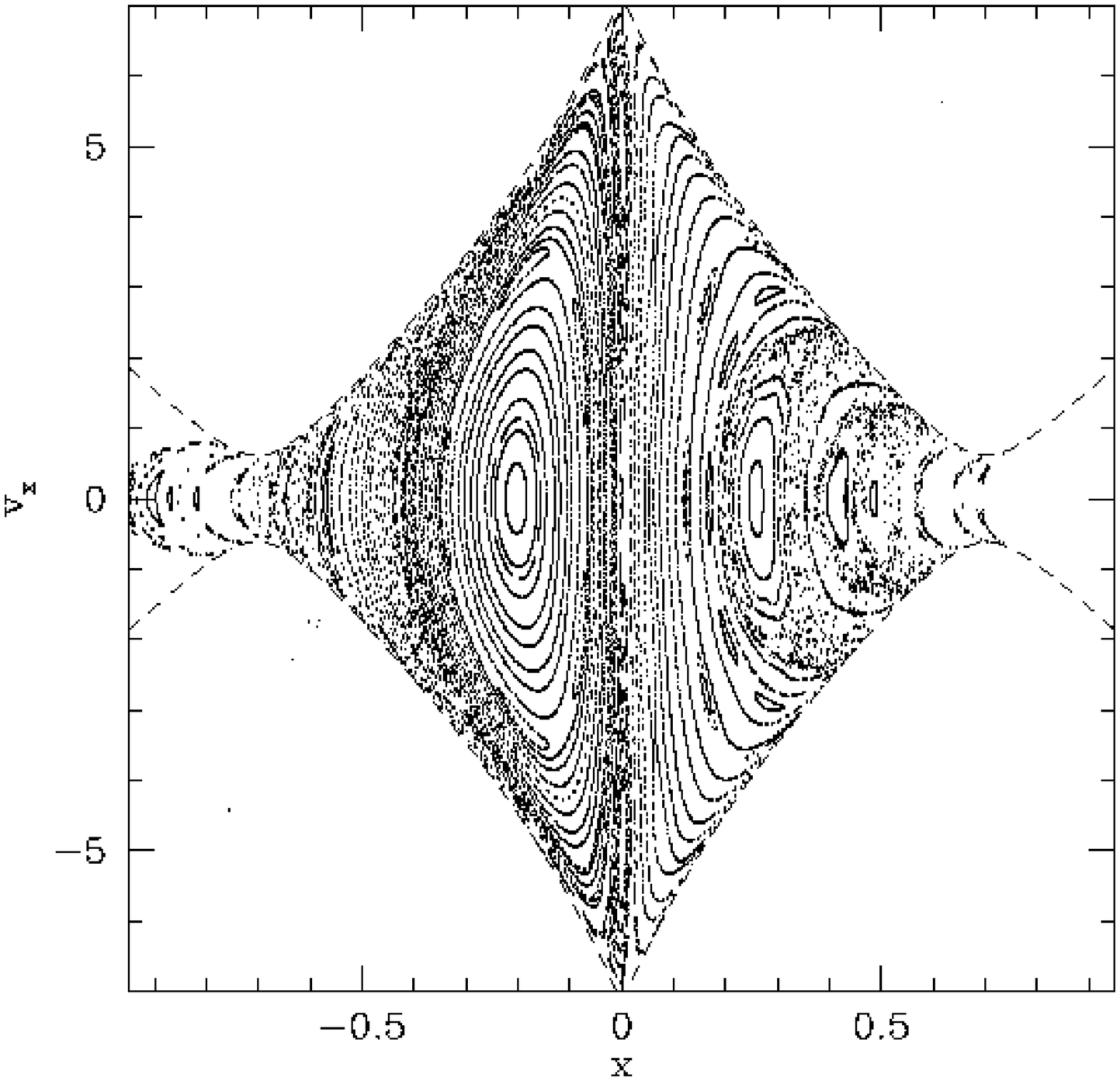}}\\
\resizebox{\hsize}{!}{\includegraphics*{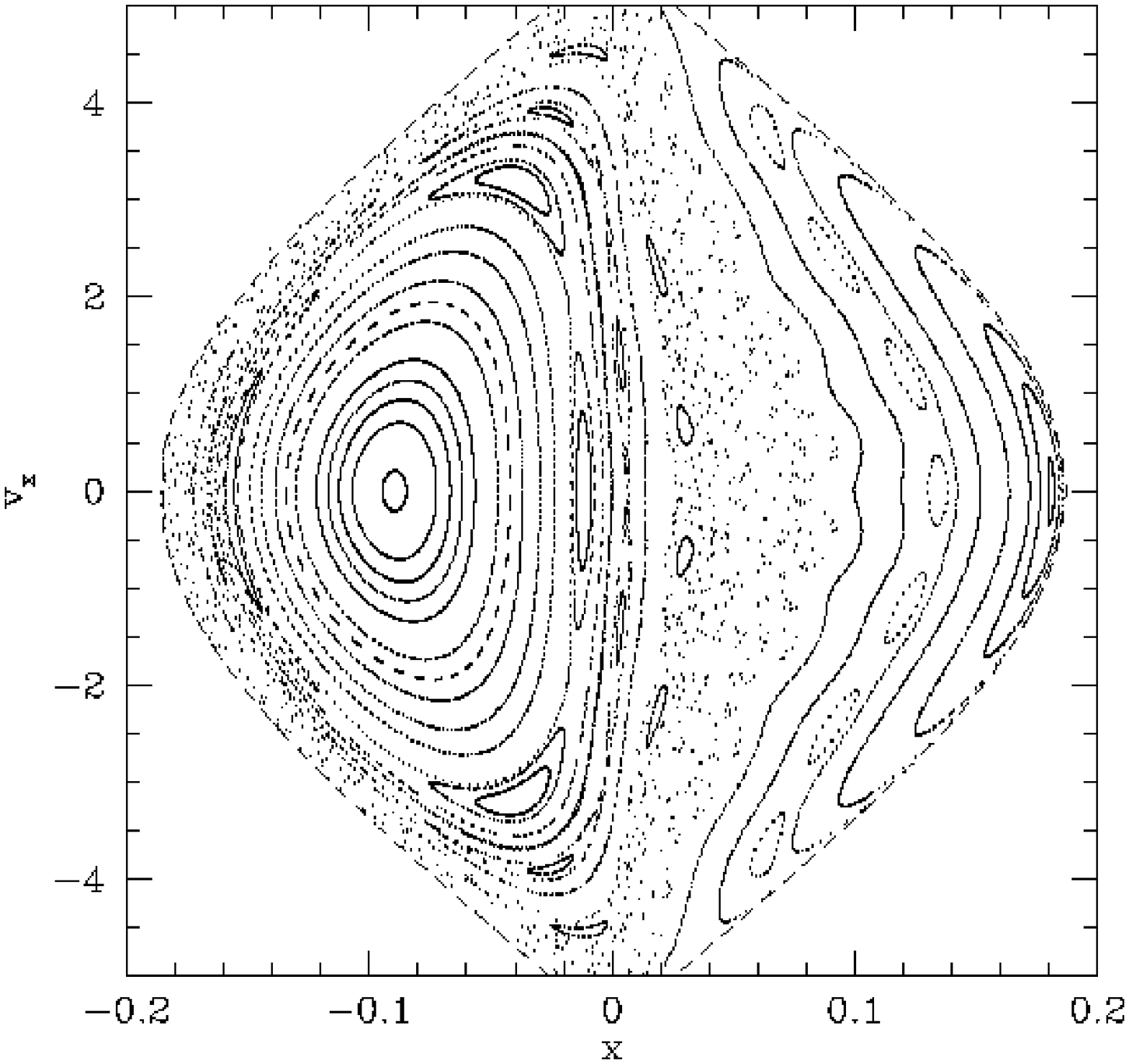}\hspace{0.1cm}
\includegraphics*{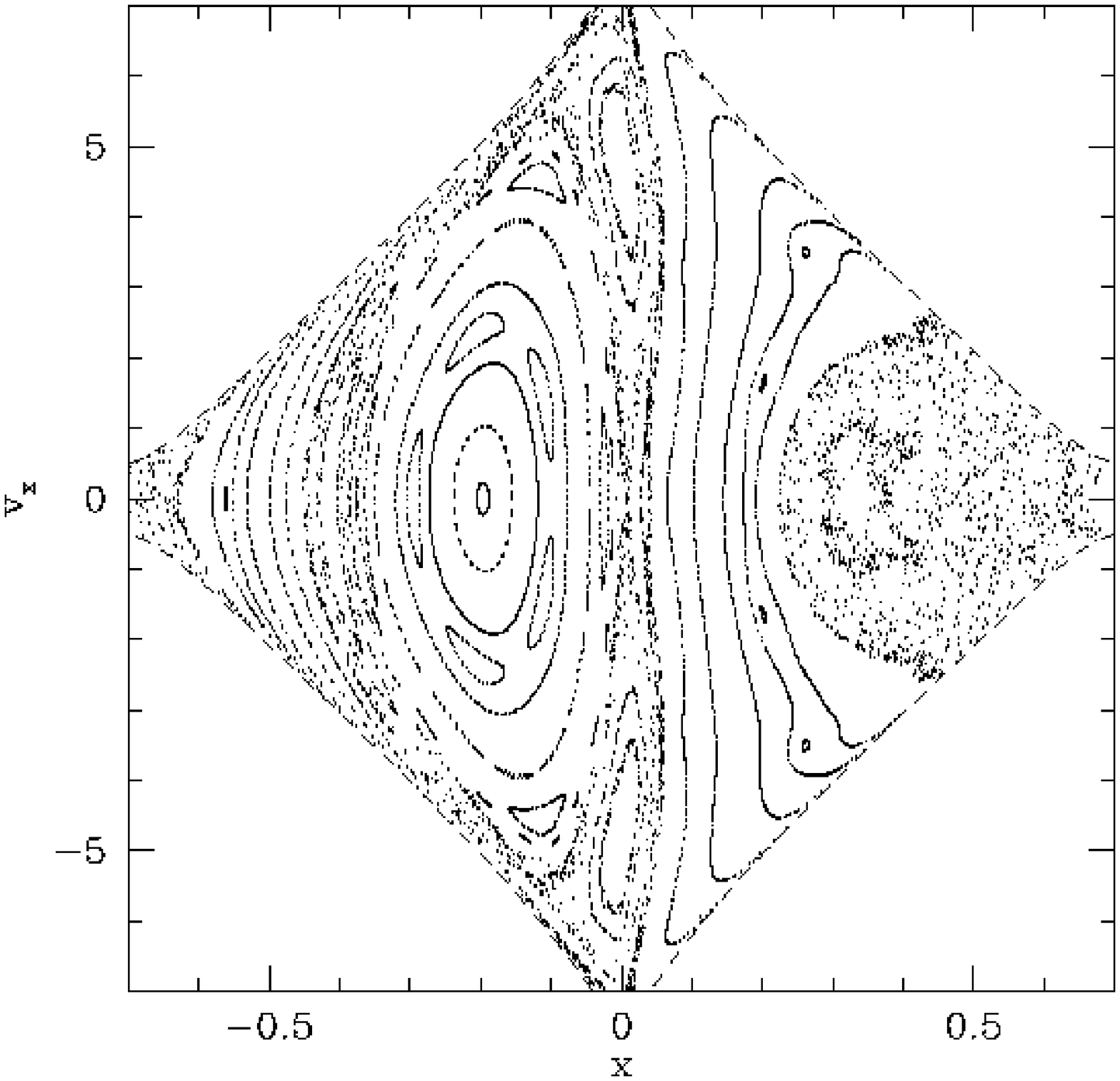}\hspace{0.1cm}\includegraphics*{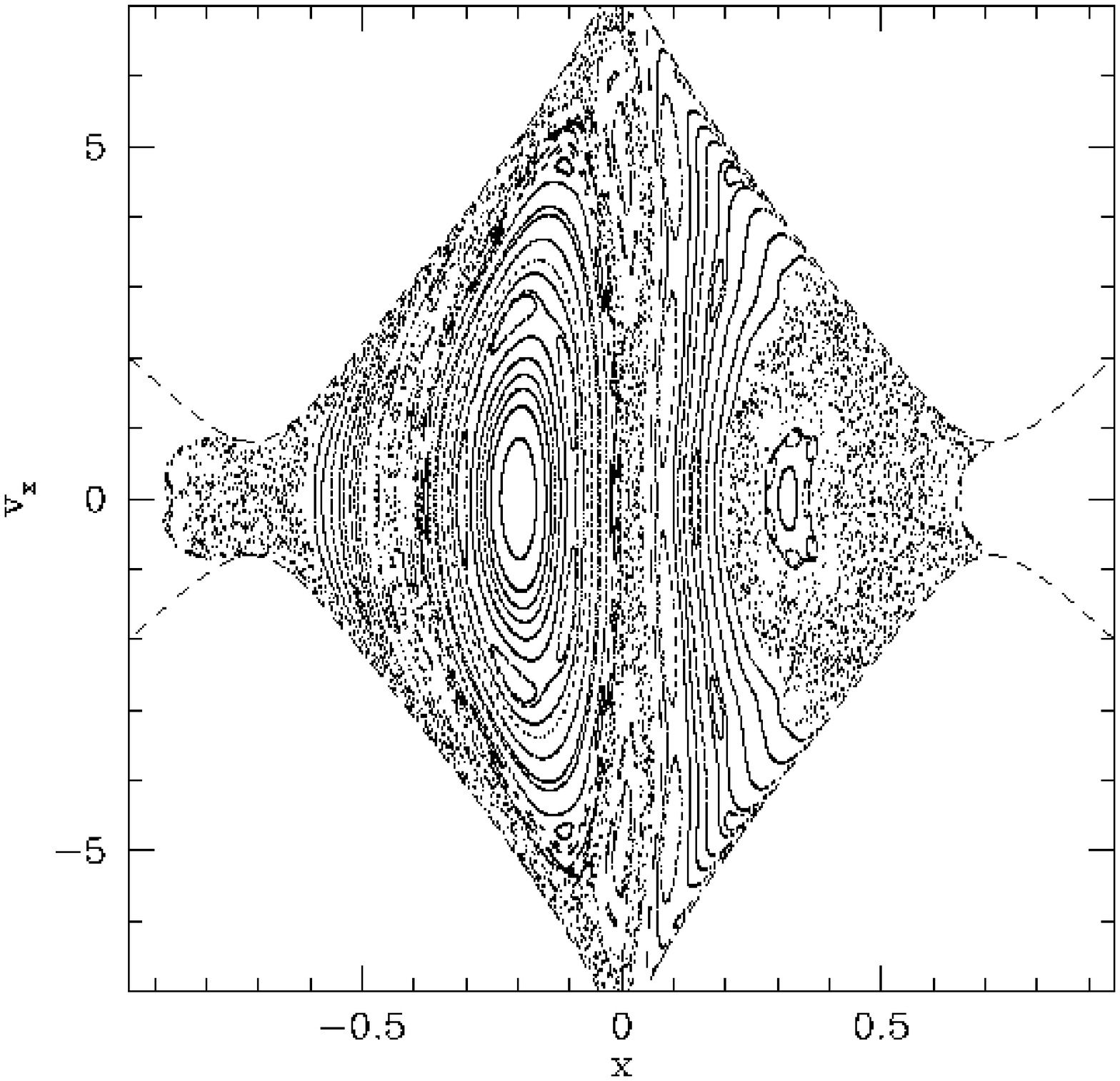}}
\caption{\em 
$SOS$s for model 1 ($C_{bar}=1$, upper row) and 2 ($C_{bar}=5.0$,
lower row).
Energy values are increasing from left to right. The values are: 
$E_J=-30.0,\,-24.8,\,-24.6$ (model 1) and $E_J=-40.0,\,-26.5,\,-26.3$ 
(model 2). Corotation is placed at $r_{CR}=0.7$.} \label{2336bm1}
\end{figure*} 
On the other hand, $\Omega_p$ has to be varied independently (with all
other parameters fixed) to study the influence of different corotation
radii $r_{CR}$ on the NGC\,2336-kinematics.  The results of those
variations are helpful for positioning $CR$ with regard to the bar
length. This is a step that can only roughly be performed using
morphological features (e.g.~bifurcations, spiral arms or rings).
\item
The profile parameter of the bulge, $\gamma$:
Different values of $\gamma$ change the radial mass
distribution of only the bulge component. The effects of such changes are
already  examined by the variations of $r_{b}$ in a sufficient
manner. Therefore, we adopt $\gamma=0.9$ for all models. 
\end{enumerate}

To illustrate the effects of the variations of the model parameters, the Surfaces of
Section and the rotation curves of four different models are shown in
the following two Sections.  Section 9.1 deals with two models with
different bar mass $M_{bar}$. In Section 9.2, the effects of two
different disk masses $M_{disk}$ are studied. Section 9.3 describes
the effects of various $\Omega_p$-values. 

\subsection{Variation of the bar mass $M_{bar}$}
\begin{figure}[t]
\resizebox{\hsize}{!}{\includegraphics*{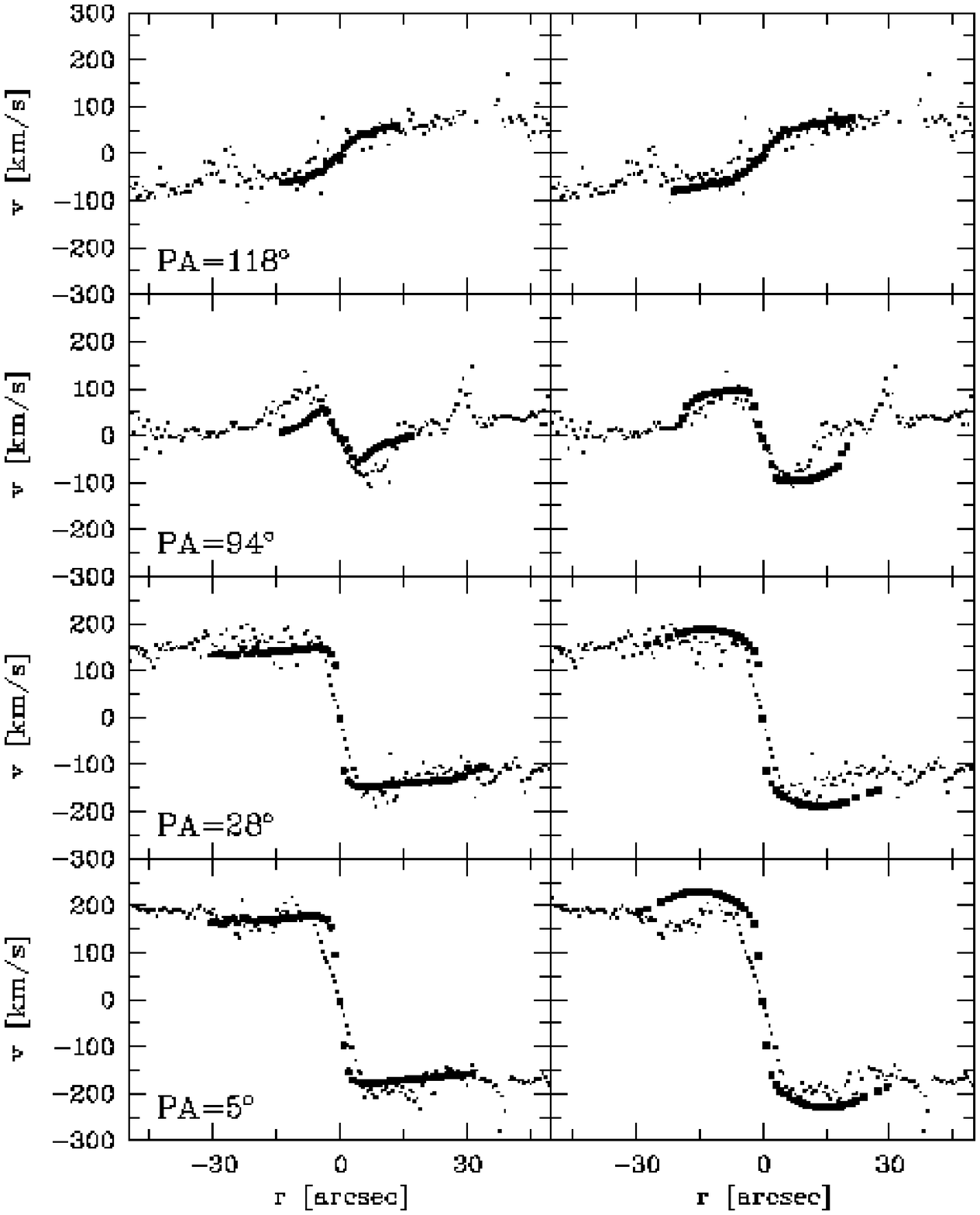}}
\caption{\em
Theoretical gas rotation curves of model 1 ($C_{bar}=1$, left column)
and model 2 ($C_{bar}=5.0$, right column). Only $x_1$-orbits are considered.
Observed points are always marked by small dots without error bars.  }
\label{2336bmrot1}
\end{figure}
The bar mass is changed by variations of $C_{bar}$.
Table \ref{modbm1} lists the basic parameters for two models
with different bar masses. Model 1 with a $M/L$ of $C_{bar}=1.0$
uses the observationally determined value.
Model 2 uses $C_{bar}=5.0$ which obviously is not confirmed observationally but
is useful for illustration purposes.  In both cases, $CR$ is placed at
the end of the bar: $r_{CR}=0.7LU$.

In general, the division of phase space resembles the one of the
introductory model: $IC$s around the periodic $x_1$-orbit are
dominating the prograde side and $x_4$-$IC$s cover the retrograde
one. $x_2$-orbits are completely absent for both models, even at very
low $E_J$-values that are not shown here.

Comparing the Surfaces of Section in Fig.~\ref{2336bm1}, a much larger
percentage of semi-ergodic motion for model 2 (lower row) can be
detected at every energy level.  This result supports a well-known
fact: massive bars with their strong non-axisymmetric potential
favour the existence of numerous semi-ergodic orbits.
In the $SOS$ of model 2 at low energies (lower row, left plot),
a small number of islands around the $IC$s of the $x_4$-family is
located within a sea of irregular or semi-ergodic motions. Due to the
limited resolution of the grid and the limited time over which the
non-periodic orbits are traced by our integration program, the space
between the $IC$s is not completely filled wih islands and single
points.

Further differences between model 1 and 2 occur when one approaches 
higher energies (central and right columns in Fig.~\ref{2336bm1}) around
$E_{CR}=-24.7$ (model 1) and $E_{CR}=-26.4$ (model 2), respectively.
The (4:1)-resonance, visible at $x\approx 0.25\ldots 0.3$
(Fig.~\ref{2336bm1}, central column), is much stronger for model 1
than for model 2.

When comparing the artificial HII-rotation curves of both
NGC\,2336-models in Fig.~\ref{2336bmrot1}, significant differences
occur for nearly all slit orientations: Model 2 (right column)
predicts higher velocities for the bar-dominated inner region of
NGC\,2336, especially for slit positions that trace the streaming
velocities of particles which are moving on $x_1$-orbits along the bar
($PA=94^{\circ}$). While model 1 (left column) seems to be a good
description for the HII-kinematics for slit orientations along the
major axis of the disk ($PA=5^{\circ}$ and $PA=28^{\circ}$), model 2
produces velocities that are too high up to regions around $r\approx
25''$. Only along the major axis of the bar
($PA=118^{\circ}$), no differences between model 1 and model 2 are visible.

Model 1 mainly suffers from the fact that the overshootings in the rotation
curves at $PA=94^{\circ}$ cannot be traced to their full extent:
velocities are too low by $\vert \Delta v_{rot}\vert\approx 45$km/sec 
when comparing the peak values of observed and predicted
rotation curves. In addition, the radial extent of these humps is
too small by a factor of 2.

A general mismatch between observations and model calculations occurs
in the innermost regions up to $r\approx 5''$, where the slopes of the
artificial rotation curves of model 1 and 2 are too steep, compared
with the observed ones.  This is a hint at a too high bulge mass
concentration in both models.

The results of this section show that $C_{bar}=1.0$ is too low
to explain the observed velocity field in
the inner regions, while
$C_{bar}=5.0$ is too high. The optimal value should lie in the range
$1.0\le C_{bar}\le 2.0$.

Second, $b/a$-ratios which are close to the photometrically derived
value ($0.3$) or agree with it, are reasonable, since it seems possible
to achieve a good agreement between the model and the observations.
If discrepancies between observed and predicted velocity field are
still remaining for observationally confirmed bar parameters, other
parameters which are closely connected to the kinematics of the inner
regions of NGC\,2336 will have to be adjusted. E.g., the velocity
contributions that are still missing in model 1 at $PA=94^{\circ}$
with reasonable bar parameters could be generated by slight changes of
$\Omega_p$ or by raising the central mass ($M_{bulge}$).

\subsection{Variation of the disk mass $M_{disk}$}
\begin{table}[t]
\begin{center}
\begin{tabular}{|l|c|c|}\hline
\bf Model &\bf 3&\bf 4\\[1.5ex]
\hline
\multicolumn{3}{|c|}{bar}\\
\hline
$a_{bar}$&0.7/7.17&0.87/7.17\\
$b_{bar}$&0.245/2.51&0.245/2.51\\
$b/a$&0.35&0.35\\
$C_{bar}$&1.5&1.5\\[1.0ex]\hline
\multicolumn{3}{|c|}{bulge}\\
\hline
$\gamma$&0.9&0.9\\
$r_b$&0.20/2.04&0.20/2.04\\
$q$&0.4&0.4\\
$C_{bulge}$&0.63&0.63\\[1.0ex]\hline
\multicolumn{3}{|c|}{disk}\\
\hline
$r_d$ &0.46/4.76&0.46/4.76\\
$C_{disk}$&5.0&2.5\\[1.0ex]\hline
$\Omega_p$ &20.3&22.2\\ \hline
\end{tabular}
\end{center}
\caption{\label{moddm1}\em
Model parameters of models 3 and 4 with different disk masses. Scale
lengths are given in model (LU) and physical units (kpc), all other
units are the same as in Table \ref{modmb1}. Notice
the changes of $\Omega_p$ that are caused by the variations of the radial
mass distribution.
} 
\end{table} 
\begin{figure*}[t]
\resizebox{\hsize}{!}{\includegraphics*{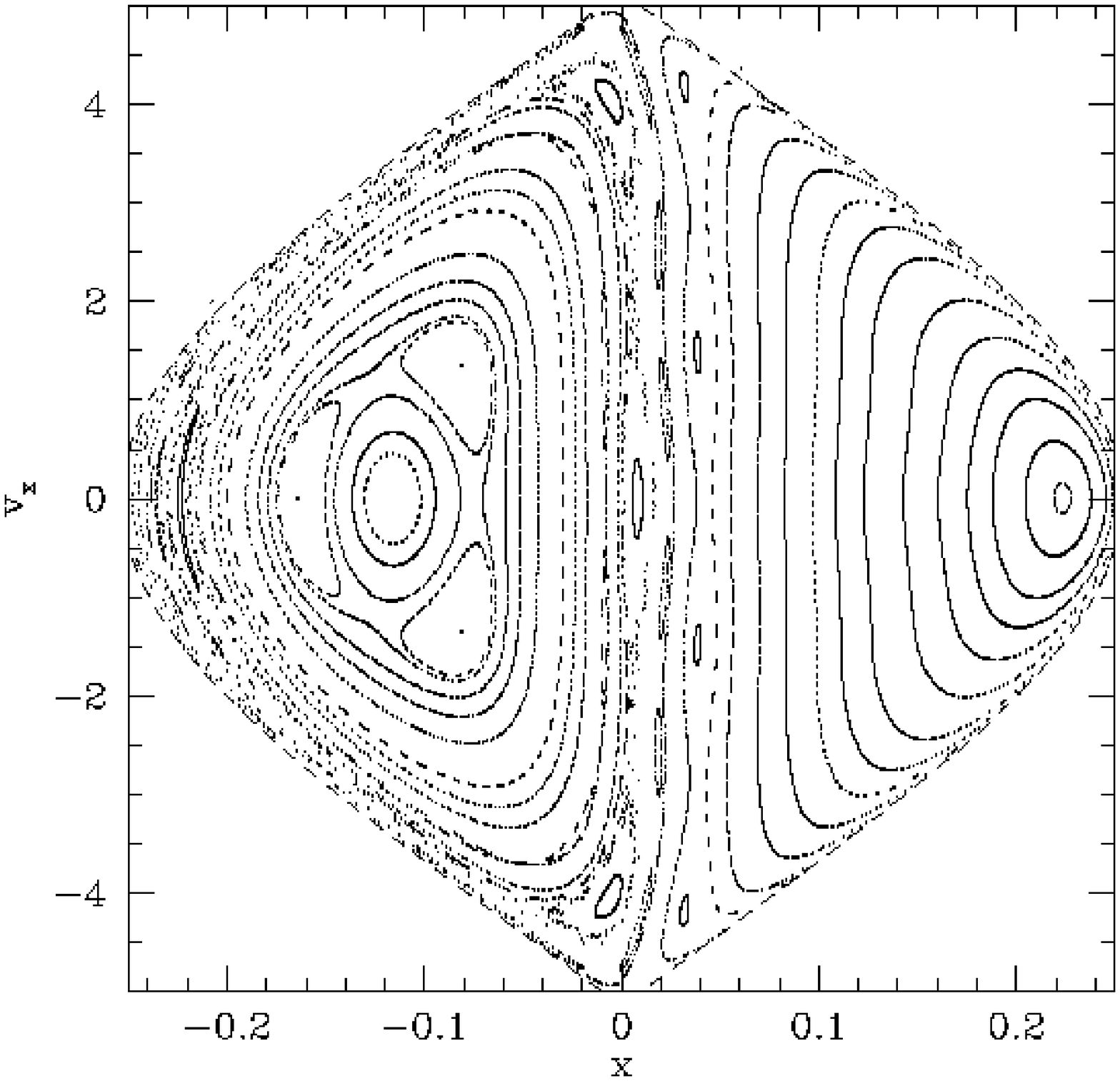}\hspace{0.1cm}
\includegraphics*{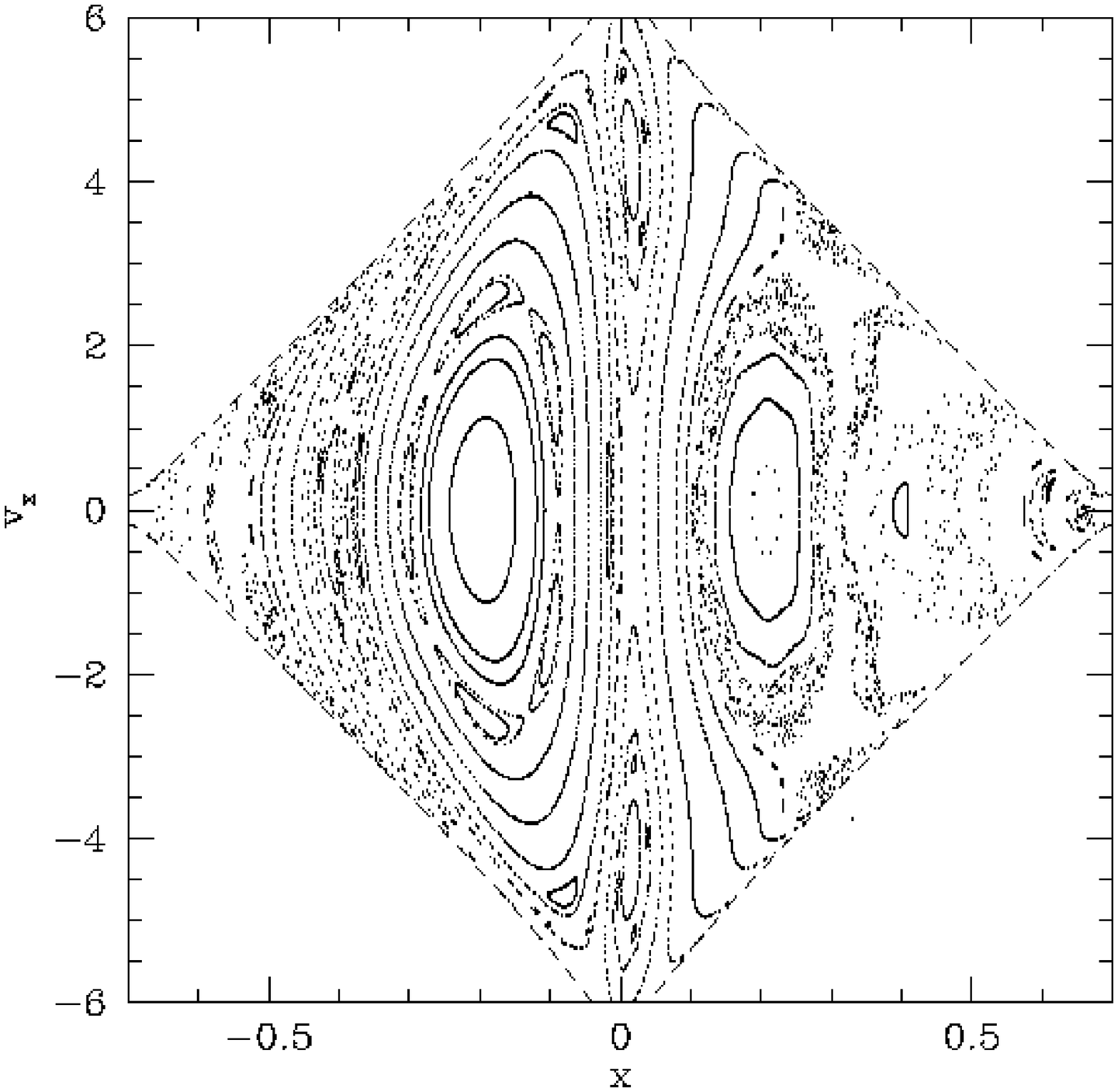}\hspace{0.1cm}\includegraphics*{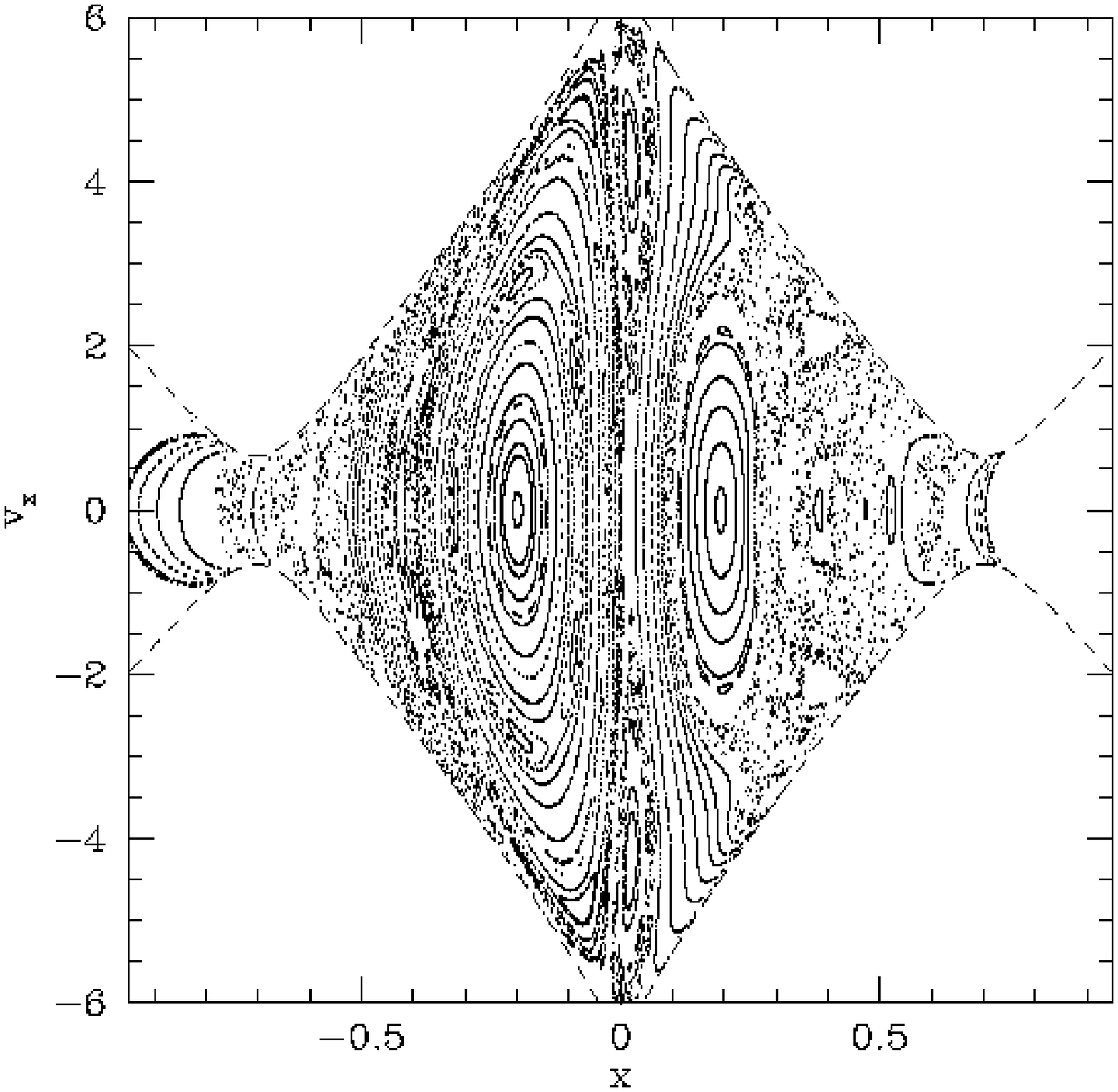}}\\
\resizebox{\hsize}{!}{\includegraphics*{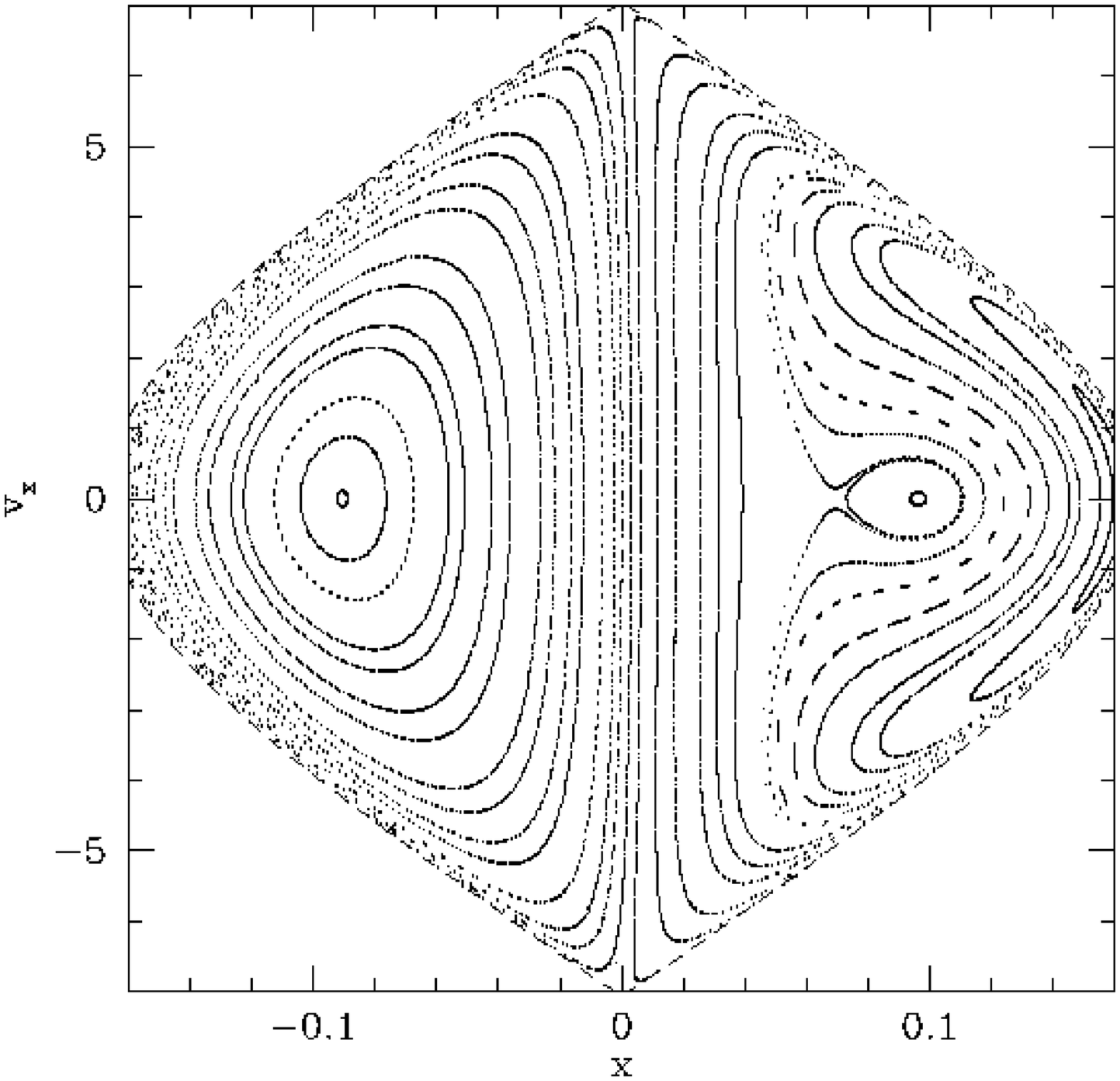}\hspace{0.1cm}
\includegraphics*{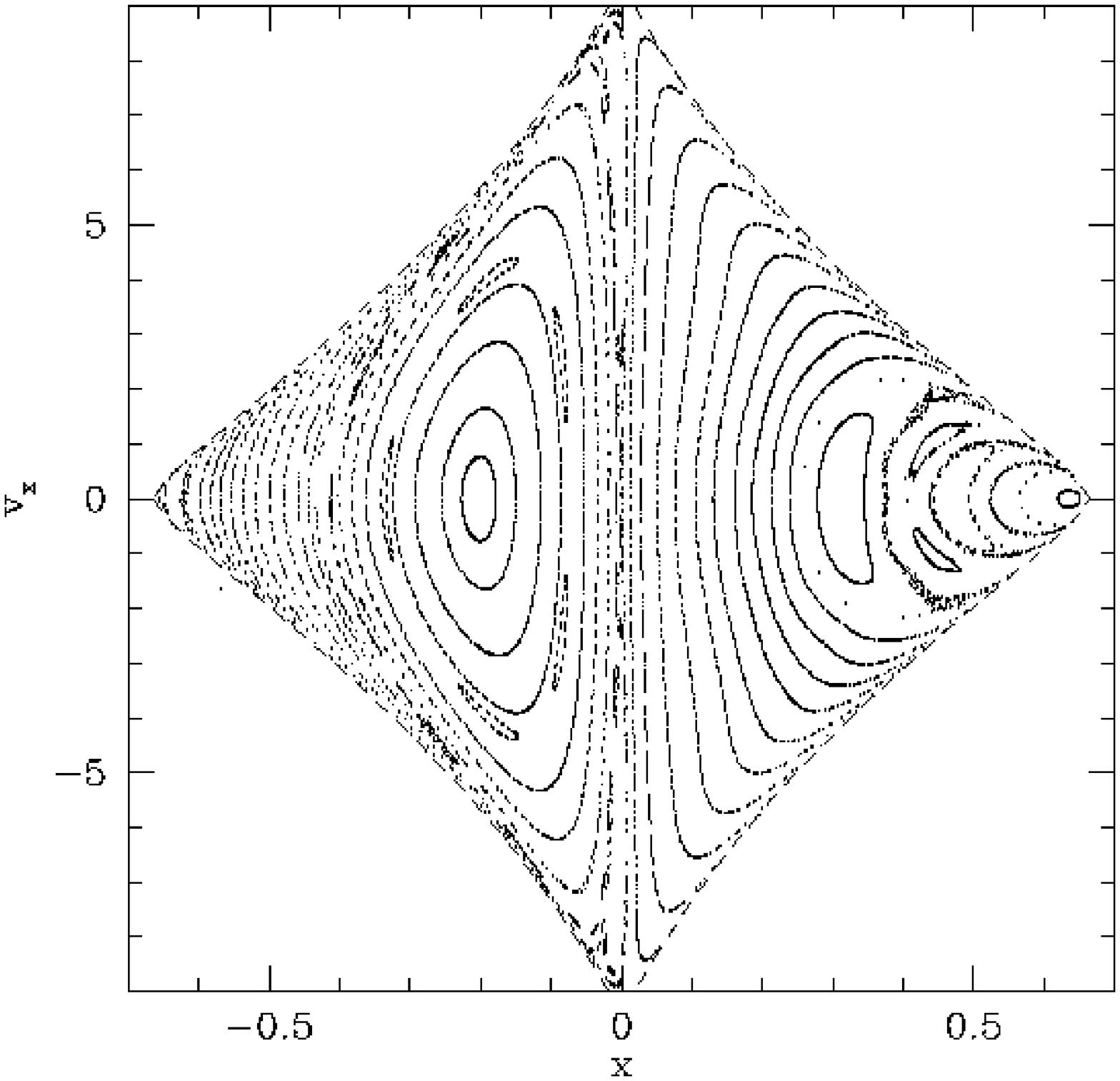}\hspace{0.1cm}\includegraphics*{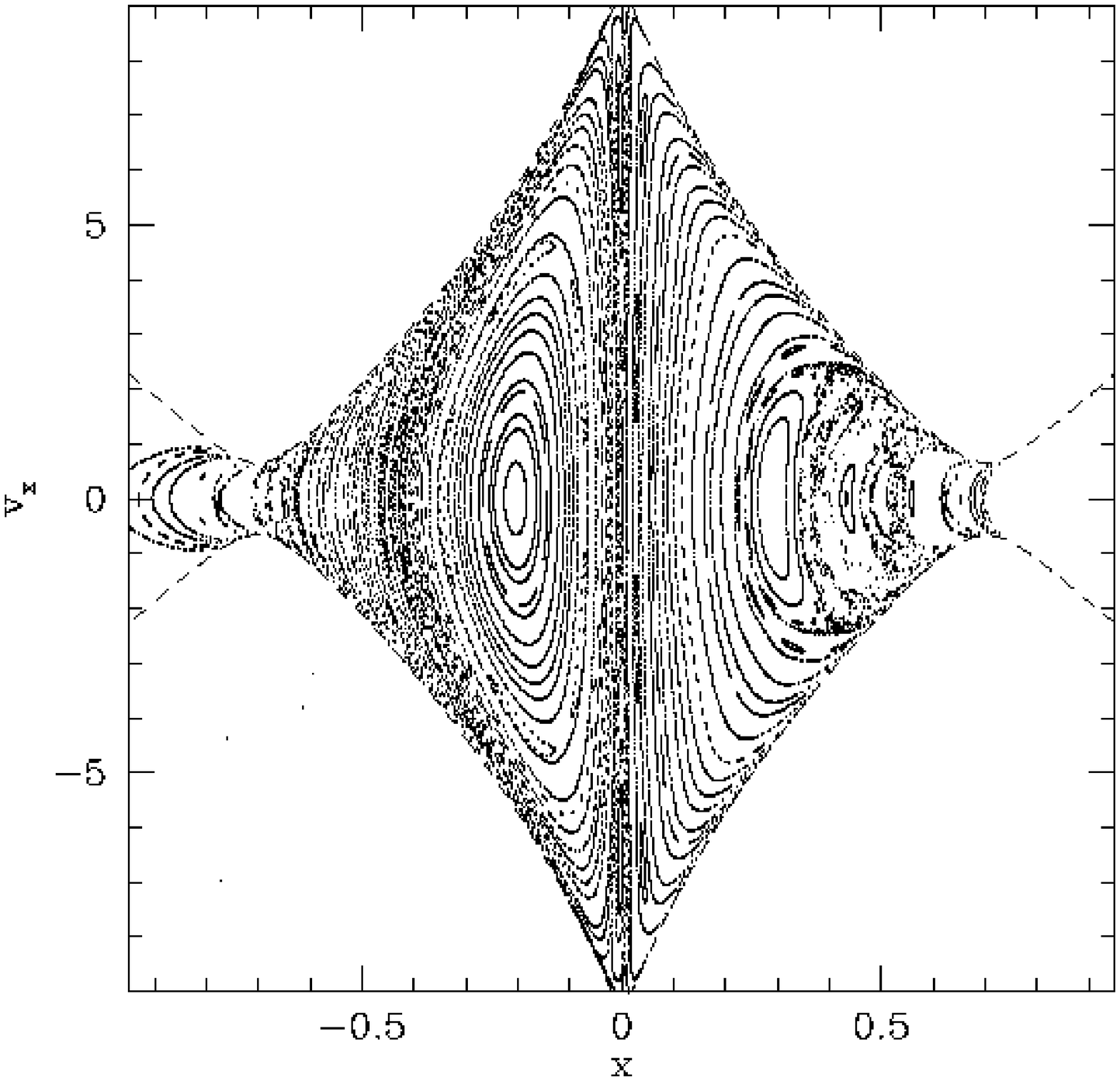}}
\caption{\em 
Surfaces of Section for model 3 ($C_{disk}=5.0$, upper row) and model
4 ($C_{disk}=2.5$, lower row). 
$E_J$-values are (from left to right): $-40.0,\,-32.4,\,-32.2$ (model 3)
and $-55.0,\,-35.8,\,-35.5$ (model 4). In both models, Corotation is
placed at $r_{CR}=1.1a_{bar}$. } 
\label{2336dm1}
\end{figure*}
To examine the effects of different disk masses $M_{disk}$,
Models 3 and
4 include disks with a $C_{disk}$ of 5.0 and 2.5, respectively. 
In contrast to models 1 and 2, $CR$ is placed at $r_{CR}=1.1a_{bar}$ in 
both cases.

\begin{figure}[h]
\resizebox{\hsize}{!}{\includegraphics*{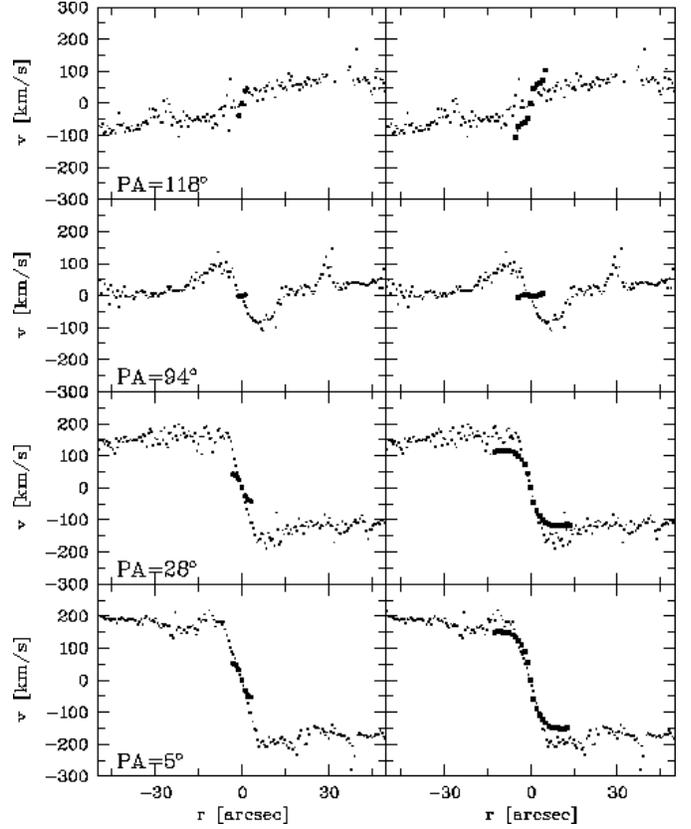}}
\caption{\em
$x_2$-rotation curves of model 3 ($C_{disk}=5.0$, left column) and 
model 4 ($C_{disk}=2.5$, right column). Observational values are marked 
by dots. While model 3 produces $x_2$-orbits for a neglegibly small 
central area of NGC\,2336 only, much more orbits are obtained in model 
4. Similarly as in models 1 and 2, $x_2$-orbits do not reproduce the observed 
velocities, either, especially for a slit orientation along the disk 
minor axis ($PA=94^{\circ}$). }
\label{2336dmrot2}
\end{figure}

The main difference between the two models is the presence of a
periodic $x_2$-orbit in model 4 at $x\approx 0.09$ (Fig. \ref{2336dm1}), surrounded by
$IC$s.  Those orbits are completely missing in the corresponding
Surface of Section of model 3 at comparable $E_J$-values. Model 3
produces $x_2$-orbits only in the very center of NGC\,2336. This fact
is illustrated by the corresponding $x_2$-rotation curves shown in
Fig.~\ref{2336dmrot2} which cover a very small region
only. Additionally, the closed $IC$s of model 3 around the $x_4$-orbit
are surrounded by islands and semi-ergodic regions already at those
low energies, a feature that is missing in model 4.

Semi-ergodicity becomes more important when we proceed to higher
energies (central and right column in Fig.~\ref{2336dm1}): At
$E_J=-32.4$, in model 3 the prograde side of the phase space is
dominated by a very strong (4:1)-resonance at $x\approx 0.2$ which is
surrounded by large regions of semi-ergodic motions (upper row,
central plot).
In contrast to this, model 4 (lower row, central plot) exhibits a much
weaker (4:1)-resonance at $x\approx 0.3$ and nearly no semi-ergodic
motion at comparable energies ($E_J=-35.8$). Even at the highest
energy values slightly above $E_{CR}$ (right column), semi-ergodic
motion plays a less important role in case of model 4 than in
model 3.

\begin{figure}[h]
\resizebox{\hsize}{!}{\includegraphics*{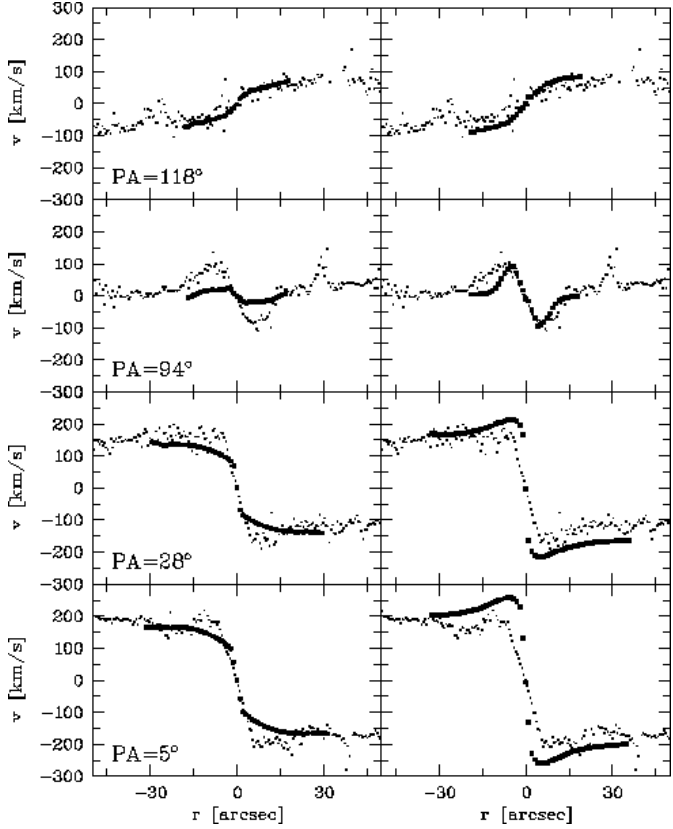}}
\caption{\em
$x_1$-rotation curves of models 3 ($C_{disk}=5.0$, left column) and 
model 4 ($C_{disk}=2.5$, right column).
Observational values are marked by
dots. Both models exhibit significant differences
at nearly all position angles.} 
\label{2336dmrot1}
\end{figure}

Both models are not able to reproduce the observed kinematics with
sufficient accuracy:
A $C_{disk}$ of $5.0$ produces $x_1$-velocities that are much too low
for the central $x_1$-orbits (model 3, Fig.~\ref{2336dmrot1}, left column).  A
moderately lower $C_{disk}$-value of $2.5$ yields a better fit (model
4, Fig.~\ref{2336dmrot1},
right column, see also the results of the optimal model in
Fig.~\ref{2336opt1} which were obtained with $C_{disk}=2.5$).
In addition, model 3 suffers from the fact that the humps and dips in
the observed HII-rotation curves at  $PA=5^{\circ},\,28^{\circ}$ and
$94^{\circ}$ are not reproduced to their full extent. This is caused
by the large overall disk mass ($C_{disk}=5.0$), with a
correspondingly small 
non-axisymmetric contribution of the bar to the total potential.
Therefore, the $x_1$-orbits supporting the bar structure become
less elongated, and the streaming velocities of the HII-clouds along the
major axis of the bar are much lower.

However, a simple reduction of $C_{disk}$ with all the other
parameters remaining fixed is no solution either, as is quite obvious
from the results of model 4 (Fig.~\ref{2336dmrot1}, right column): With
$C_{disk}=2.5$, the central slopes of the rotation curves of this model
are too steep due to the increased importance to the non-axisymmetric
bar contribution to $\Phi_{total}$ and to the small bulge scale length
$r_b$.  Additionally, the maximum velocities of model 4 exceed the
observed ones by $\vert\Delta v_{rot}\vert \approx 50-100$km/sec at
$PA=5^{\circ}$ and $PA=28^{\circ}$. Another feature of model 4 is the
strong decline of the rotation curves after reaching the (correct)
maximum velocities for a slit orientation of $PA=94^{\circ}$, which is
not supported by observations, either.

It can be shown that a slightly
larger bulge mass ($C_{bulge}\approx 1.0$ instead of
$C_{bulge}=0.64$), increasing
the mass concentrated in the center of NGC\,2336, extends the HII-streaming
motions along the bar to larger radii ($PA=94^{\circ}$).
At the same time, a larger $r_b$-value would induce a
flatter rise of the velocities in the innermost part of NGC\,2336 and
reduce the observed humps in model rotation curves at $PA=5^{\circ}$
and $28^{\circ}$ to the correct height.

Both models produce $x_2$-orbits, the corresponding rotation curves 
are also computed and shown in Fig.~\ref{2336dmrot2}. While model 3 
(left column) generates $x_2$-orbits only in the very central
region ($\le 3''$) of NGC\,2336, in model 4 (right column) $x_2$-orbits are 
present up to radii of $r\approx 15''$ in the case of $PA=5^{\circ}$ and
$28^{\circ}$.  As far as those orbits are present in both models, no
significant differences occur between the $x_2$-rotation curves.
However, supporting the results of models 1 and 2, the $x_2$-orbits of models 3
and 4 cannot explain the kinematics of the inner regions of NGC\,2336,
which is clearly visible for a slit position of $PA=94^{\circ}$.
To summarize, large $C_{disk}$-values lead to an increased
axisymmetric potential component, therefore the bar becomes less
important. The consequences of the massive disk in model 3 are
$x_1$-streaming velocities that are in general too low by far to
reproduce the observed velocity field.
We draw the conclusion that a
final model of NGC\,2336 should use a $C_{disk}\le 3$. The necessary
adjustment of the mass distribution in the central region could then
be achieved by varying other parameters, e.g.~the bulge mass, which
was chosen especially low for the models 3 and 4. The slope of the
rotation curves could be adjusted by varying $r_b$, the streaming
velocities by lowering $\Omega_p$.

\subsection{Variation of the bar pattern speed $\Omega_p$}
In contrast to the parameters resulting from the morphological
decomposition (e.g.~component masses and scale lengths), $\Omega_p$
cannot be observed directly. Therefore a sequence of identical models
with different $\Omega_p$-values is examined.
In general, all models produce similar velocity fields. Major
differences occur only for a slit orientation of $PA=94^{\circ}$,
where models with lower $\Omega_p$ predict significantly larger
humps in the rotation curves.
This is a consequence of the fact that the rotation curves result from
the $x_1$-streaming velocities -- which are identical for all models --
subtracted by the various $\Omega_p$-values.
In contrast to that, the other slit
orientations show only minor deviations.
Obviously, variations of $\Omega_p$ are an appropriate tool to adjust
the velocities perpendicular to the bar major axis.

Upper and lower limits for $\Omega_p$ are given by the following
arguments: Since $r_{CR}$ must be located outside the bar, $\Omega_p$
must not exceed a certain value, depending on the mass distribution in
the model. This leads to an upper limit for $\Omega_p$.
Estimating the lower limit is more complicated since the height of the
humps in the rotation curves is not exclusively determined by
$\Omega_p$, but is also influenced by several other model parameters. 

Examples for this mutual influence of several parameters are the cases
of models 1 and 2:
Numerical N-body-simulations
of barred galaxies strongly suggest that the $CR$ of intermediate type
galaxies (e.g.~NGC\,2336) normally is located at
$r_{CR}\approx 1.1a_{bar}$.
In models 1 and 2 (Section 9.1), $CR$ is placed at the end of the bar at
$r_{CR}=1.0a_{bar}$,
which in
fact is the lower boundary for $r_{CR}$.
Even for the more realistic model 1 with the lower bar mass, the
HII-velocities are still too small. 
Therefore, if we reduce $\Omega_p$ of model 1 to a value
shifting Corotation to $r_{CR}=1.1a_{bar}$, the velocity humps of the 
$PA=94^{\circ}$-rotation curve increase towards values that are in much better
agreement with the observations.

\section{The Optimal Model for NGC\,2336}
In the last Section we used models with different bar masses (1 and 2)
and different disk masses (3 and 4) to illustrate the qualitative and
quantitative changes in the division of the phase space and the
HII-kinematics which are caused by the variation of $M_{bar}$ and
$M_{disk}$. The effect of $\Omega_p$-variations was described as well.
But as was discussed in the beginning of Section 9, there are many more
parameters which may induce changes in the predicted kinematics,
e.g.~the scale lengths of disk, bulge and bar, $r_d$, $r_b$, and $a$. 
The results of the studies of
changes in the latter parameters, as well as the results of the models
1-4, can be summarized in the following constraints applying to the
construction of the final NGC\,2336-model:
\begin{enumerate}
\item
Since changing the vertical flattening of the bulge, $q$, results in
minor changes in the HII-rotation curves only, $q=0.4$ is chosen for
the optimal model, which is well in agreement with the photometric
observations.
\item
The examination of models with different offset angles $\psi$
between the bar and the $LON$ shows that a deviation of a few degrees
($\pm 2^{\circ}$) leads to small changes in the model gas
kinematics that tend to change the rotation curves in the same way as
do variations of the bar axis ratio $b/a$.  Since the morphological
decomposition of NGC\,2336 suggests $\psi=104^{\circ}$ as the most
probable number and since a good fit to the observed velocity field
can be achieved with this value, $\psi=104^{\circ}$ will be used for
the final model.
\item
Models with $\Omega_p$ being varied support those values that place $CR$ at
$r_{CR}=1.1a_{bar}$. Placing $r_{CR}$ outside the bar in an interval
$r_{CR}=1.0\ldots 1.4a_{bar}$ is supported by results from numerical studies
(cf.~Athanassoula~\cite{athanassoula} and references herein). 
\item
A bar-$b/a$ of $0.3$ is in good agreement with the kinematic 
requirements and the observations as well.
\item
The bar mass-to-light ratio should not exceed $C_{bar}=2.0$.
\item
Due to the difficulties in determining the bulge scale length, $r_b$,
this parameter cannot be determined exactly from observations. The
results of the wide range of models examined suggests
that $r_b$ should lie within $r_b=0.2\ldots 0.3LU$ ($2.1\ldots 3.1$kpc). 
\item
The photometrically derived bulge mass ($C_{bulge}=1.0$) is a
reasonably good choice.
\item
The observed disk scale length of $r_d=0.46LU$ ($4.76$kpc, unchanged in all models)
agrees well with the requirements of the rotation 
curves, assumed that all other parameters are precisely adjusted as well. 
\item
With $r_d$ unchanged, a small disk mass (i.e.~$C_{disk}\le 3.0$) seems
to be a plausible  choice when one tries to fit the model rotation
curves to the observed kinematics in the outer disk regions at $r\ge
50''$ via adjusting the cloud speeds on the $x_1$-orbits beyond $CR$.
\item
In order to adjust the height of the overshootings in the rotation curves for a
slit orientation $PA=94^{\circ}$ and to place $CR$ at
$r_{CR}=1.1a_{bar}$, 
the pattern speed $\Omega_p$ has to take low values $\le 20$km/sec/kpc.
\end{enumerate}
The final model was optimized by studying 32 parameter variations with
total number of $\approx 9500$ closed orbits.  Finally, the best results were
obtained with the parameter set listed in Table \ref{modopt1}.
\begin{table}[h]
\begin{center}
\begin{tabular}{|l|c|}\hline
\multicolumn{2}{|c|}{\bf bar}\\[0.3ex]\hline
$a_{bar}$ [LU/kpc]&$0.7/7.17$\\
$b_{bar}$ [LU/kpc]&$0.21/2.15$\\
$b/a$&$0.3$\\
$M_{bar}$ [$M_{\sun}$]&$1.13\cdot 10^{10}$\\
$C_{bar}$& 1.5\\[1.0ex]
\hline
\multicolumn{2}{|c|}{\bf bulge}\\[0.3ex]\hline
$\gamma$&$0.9$\\
$r_b$ [LU/kpc]&$0.23/2.36$\\
$q$&$0.4$\\
$M_{bulge}$ [$M_{\sun}$]&$1.20\cdot 10^{10}$\\
$C_{bulge}$& 1.0\\[1.0ex]
\hline
\multicolumn{2}{|c|}{\bf disk}\\[0.3ex]\hline
$r_d$ [LU/kpc]&$0.46/4.76$\\
$M_{disk}$ [$M_{\sun}$]&$9.63\cdot 10^{10}$\\
$C_{disk}$& 2.5\\[1.0ex]\hline
$\Omega_p$ [km/sec/kpc]&$16.40$\\ \hline
\end{tabular}
\end{center}
\caption{\label{modopt1}\em
Parameter values for the optimal NGC\,2336-model. All scale lengths
are given in model (LU) and physical units (kpc), using a distance of
$22.9$ Mpc for NGC\,2336.
}
\end{table}

The
total mass of the final model was calibrated according to equation 28
and equation 29 with the outer circular velocity value from the
kinematical observations. We obtain a total mass of $M_{tot}=1.20\cdot
10^{11}M_{\odot}$. If we identify for a rough estimate the relative
mass-to-light-ratios $C_{disk}$ and $C_{bar}$ with the normal
astrophysical values, we obtain from our optimal model:
\begin{eqnarray}
M_{bulge}&=&1.20\cdot 10^{10}M_{\odot}\\
M_{disk}&=&9.63\cdot 10^{10}M_{\odot}\\
M_{bar}&=&1.13\cdot 10^{10}M_{\odot}.
\end{eqnarray}
Thus we have a bulge/disk ratio of $B/D=0.125$ and a bar/disk ratio of
$Bar/D=0.117$. One should keep in mind that especially the disk mass
contains a certain contribution of dark matter which can not be
separated within the framework of this paper.

\begin{figure}[t]
\resizebox{\hsize}{!}{\includegraphics*{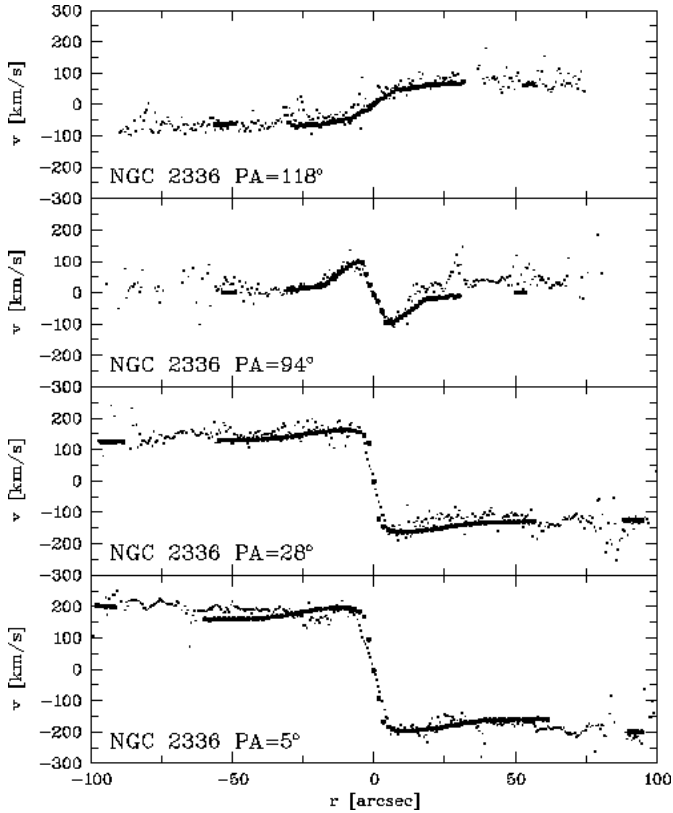}}
\caption{\em
Observed (points) and artificial rotation curves (filled circles) for
NGC\,2336. Longslit orientations are indicated according to the
position angles ($PA$) of the spectrograph. The points at large radii
after the gap are produced by the nearly circular $x_1$-orbits outside
the $CR$ radius.
}
\label{2336opt1}
\end{figure}

Looking at the artificial rotation curves in Fig. \ref{2336opt1} one
recognizes the excellent agreement between the artificial curves and
the observed ones. Not only does the optimal model produce the correct
maximum rotational velocities of the slit orientations $PA=5^{\circ}$
and $28^{\circ}$, but also traces the overshootings of the
$PA=94^{\circ}$-rotation curve correctly.  The positions of the
velocity maxima and minima in the model rotation curves and the
observed ones are identical.  The theoretical rotation curves in
Fig.~\ref{2336opt1} display gaps in the vicinity of $r_{CR}$, since
the orbit integrator can not compute orbits near the Lindblad
resonances.

Small discrepancies that remain between observation and
theory are the central parts of the rotation curves for $PA=5{\circ}$
and $PA=28^{\circ}$, where the theoretical slopes rise slightly faster
than the observed ones. 

\section{Discussion}
\subsection{Morphological Decomposition}
The morphological decomposition of the $J$-band image of NGC\,2336
seems well established, since nearly all parameters that were
initially derived from the multi-component 2D model of the $J$-band
surface brightness distribution appear in the final model with the
same numerical values.  There are justified exceptions, e.g.~the bulge
scale length $r_b$, where the limited spatial resolution of the images
and the small area covered by the bulge component prevent us from a
more precise determination. The moderate increase of the final value
of $r_b$ -- compared to the morphological decomposition --
nevertheless agrees well with observations. It is also remarkable that
the bar-$b/a$ determined from observations is nearly the same as that
used for the optimal model. This is not self-evident, as the inclined
position of the bar in the sky makes the determination of the
deprojected bar-$b/a$ difficult.

Uncertainties of the model parameters become large wherever the
spatial resolution or the area of the $NIR$-detector is concerned: The
determination of the bulge scale length $r_b$ is certainly precise to
a $50\%$ level only, due to the small bulge area (i.e.~, the small
number of pixels) and the residuals remaining after the subtraction of
the bulge and disk components.  For the disk, similar uncertainties
are introduced by the limited size of the detector chip which
omits the outer parts of the disk, leading to an uncertainty in $r_d$
of $\approx 15\%$.  The luminous masses of disk, bulge, and bar can be
determined more precisely, especially the contribution of the bar is
uncertain by $5-10\%$ only, since the very regular disk structure ensures a
reliable decomposition. For the photometrically derived values of
$M_{disk}$ and $M_{bulge}$, a precision of $15\%$ seems a realistic
estimate, due to the reasons discussed above.

The deprojection angles $\phi$, $\theta$, and $\psi$
are reliably determined with an error of $\pm 2^{\circ}$
at most, since
otherwise a mismatch between the observed kinematics and that of the
optimal NGC\,2336-model would occur in at least {\em one} slit
orientation. Obviously this is not the case.

\subsection{Analysis of Phase Space}
From the $SOS$-analysis we conclude that the best NGC\,2336-models
show a phase space partition characteristic for barred galaxies with a
bar contributing roughly $10\%$ to the total potential: For all
realistic parameter sets, an $ILR$ is produced together with
$x_2$-orbits that dominate the motion in the innermost parts of the
NGC\,2336-models at low $E_J$-values.

For energy values $E_J >E_{ILR}$, the $x_2$-orbits disappear, only the
prograde $x_1$- and the retrograde $x_4$-family remain. In the
vicinity of $E_{CR}$, the $x_1$-family dissolves and is replaced by
higher resonances, e.g.~the (4:1)- and the (6:1)-family. The
(4:1)-resonance ist not very strong, so we conclude that in time
dependent models the stellar bar of NGC\,2336 would be moderately box-shaped
only. The small amount of semi-ergodicity leads to the conclusion that
a time-dependent model with our optimal parameter set would be
secularly stable on time scales of the bar rotation period.

Summarizing the results of phase space analysis, the optimal model for
NGC\,2336 exhibits the typical behaviour of disk-dominated barred
galaxies with the bar-$b/a$ and $\Phi_{bar}$ being moderately small.

\subsection{Rotation curves}
In our studies, special emphasis was put on the reproduction of the
gas kinematics in those regions influenced by the bar, i.e.~the large
streaming motions of HII-clouds orbiting on $x_1$-orbits around the
center of NGC\,2336. However, the overall kinematics with the observed
rotational velocities in the outer regions of the disk had to be
described correctly as well.  Our final model fulfills both
requirements, as is obvious from the excellent agreement between
observed and artificial rotation curves in every part of NGC\,2336 and
for every slit orientation.

Moreover, the success of adjusting the outer parts of the model
rotation curves is an evidence for the good deprojection performed
here. Otherwise the final velocities in the outer disk regions could
not be fitted simultaneously for all slit directions.

In general, all characteristic features of all velocity curves are
traced well by the final model, especially the overshootings that can
be observed in slit orientations of $PA=5^{\circ}$, $28^{\circ}$ and
$94^{\circ}$, are produced by the model to full extent. Small
differences between observed and artificial gas velocities remain in
the innermost region of NGC\,2336 and are probably caused by the
uncertainties in the bulge scale length $r_b$ due to possible errors
in the morphological decomposition.  The precise determination of the
latter is difficult due to the limited spatial resolution of the
$NIR$-detector and the small area covered by the bulge component.

From the observational point of view there is no necessity for
including $x_2$-orbits in the final sets of closed orbits, because a
perfect fit to the observed kinematics is already achieved by
$x_1$-orbits alone. Moreover, the possibility of $x_2$-orbits
contributing significantly to the HII-kinematics in the inner part of
the bar are definitely ruled out, since no model with $x_2$-orbits can
account for the streaming velocities measured at a slit orientation of
$PA=94^{\circ}$.  Though our final model produces a moderately strong
$ILR$ and therefore numerous $x_2$-orbits, this is no contradiction to
their uselessness in the case of NGC\,2336, because our model does not
make any statements with regard to the occupation number of single
orbit families.

To correct for the presence of dark matter in the outer regions of
NGC\,2336, moderate corrections of the conversion factors $C_{disk}$
and $C_{bar}$ had to be applied in order (i) to produce the large
streaming velocities along $a_{bar}$ and (ii) to normalize the
circular $x_1$-velocities in the outer disk regions:
$C_{disk}$ changed from the observationally determined value of $1.0$
to $2.5$, while the visible bar mass ($C_{bar,\,NIR}=1.0$) had
to be increased by only $50\%$. The value $C_{bar}=1.5$ is a further hint
at a reliable morphological decomposition: For the bar component
which is not strongly affected by a possible dark matter halo, a
moderate increase of the $NIR$-$C_{bar}$ seems plausible, especially
when considering possible stellar population changes between bulge and
bar that are not known in detail.

Summarizing the results of our kinematical analysis, the final
NGC\,2336 model rotation curves are a very good representation of the
observed velocity field. Obviously, there is no necessity of including
additional hydrodynamical or magnetical effects in order to explain
the observed kinematics. It is also remarkable that these results can
be achieved based on models that neglect spiral arms and an explicit
dark halo component with independent scale length.

\section{Summary}
In this paper, 2D models of the observed $J$-band luminosity of
NGC\,2336 were constructed by fitting a disk, a bulge, and a bar to
the observed surface brightness distribution. The resulting model was
deprojected and converted to an underlying mass distribution.  The
total 2D potential, $\Phi_{total}$, was obtained by numerical
expansions of the potential of the single components and coadding
them. For the examination of test particle motions in the potential of
NGC\,2336, we used a numerical orbit integrator, in which
$\Phi_{total}$ and its derivatives are implemented. The time-dependent
motions of single HII-clouds are traced by a grid-based integration
scheme, in which particles move on arbitrary orbits in the stationary
potential of the rotating bar.

The resulting partition of the phase space was examined by analysing
the appropriate cuts through phase space at a given energy, the
Poincar\'e Surfaces of Section. They provided us with the start values
-- coordinates and velocities -- of the main orbit families
constituting the bar. Given those initial values, complete sets of
closed orbits in a certain energy range were computed and projected to
the sky. Using a virtual longslit in various orientations, artificial
rotation curves were constructed and compared to the observed
kinematics of the HII-gas.

In a further step, the effects of parameter variation on the division
of phase space and the predicted HII-rotation curves of numerous
NGC\,2336-models were examined. The effects of varied $M_{bar}$- and
$M_{disk}$-values were shown here in greater detail, while the effects
of all other free parameters being changed and studied are described
qualitatively. The models led to narrow limits for the parameter
values of an optimal model which should (i) reproduce the observed
HII-kinematics directly and (ii) use the parameter values directly
obtained from the morphological decomposition procedure wherever
possible.

From the final NGC\,2336-model, the following results are obtained:
\begin{enumerate}
\item
The model parameter values obtained from observations and those needed
for an optimal fit of the HII-kinematics are identical (with justified
exceptions), therefore the aim of constructing a consistent
NGC\,2336-model is achieved.
\item
The overall quality of the morphological decomposition model is
sufficiently good to produce a realistic mass model of
NGC\,2336. Nevertheless, to check whether the kinematically supported
scale length values of bulge and disk are fully consistent with the
more uncertain morphological ones, we would need $NIR$-data with (i) a
better spatial resolution and (ii) larger areas covered by the
detector chip. With the small pixel number and the comparably large
pixel size of the $MAGIC$-detector, the determination of $r_d$ and
especially $r_b$ remains crucial.
\item
The phase space of all acceptable models, especially the optimal one,
exhibits a structure which is typical for a disk-dominated galaxy with
a bar roughly contributing $10\%$ to the total potential, i.e.~small
amounts of semi-ergodicity as well as the presence of the $x_1$-,
$x_2$- and $x_4$-family of periodic orbits.
\item
$x_2$-orbits can definitely be excluded when one tries to fit the
observed velocities within the bar, since the streaming velocities
produced by them are by far too low.
\item
The optimal model favours a low pattern speed ($\Omega_p\approx
16$km/sec/kpc), a small bar axis ratio ($b/a=0.3$) and -- compared
with the luminous matter -- moderately increased bar and
disk masses due to the presence of dark matter ($M_{bar}=1.13\cdot
10^{10}M_{\odot}$, $M_{disk}=9.63\cdot 10^{10}M_{\odot}$).
\item
An excellent agreement between the observed and the artificial
rotation curves is achieved, the optimal NGC\,2336-model describes
correctly the large streaming motions of the HII-clouds along the bar
as well as the rotation velocities in the outer regions of the disk.
\item
In general, the case of NGC\,2336 shows that the peculiar
HII-kinematics of a strongly barred galaxy can be completely explained
by consistent stationary models which are constructed from
observationally derived parameter sets.
\end{enumerate}

\begin{acknowledgements}
The authors want to thank R.~Bender (Universit\"atssternwarte
M\"unchen) for providing the software for the Fourier quotient
correlation analysis of the stellar spectra, FCQ4. We also acknowledge
the help of J.~Heidt with the $NIR$ data reduction.  This work has
been supported by the DFG Sonderforschungsbereich 328.
\end{acknowledgements}


\end{document}